\documentclass[10pt,journal]{IEEEtran}

\newtheorem{lemma}{Lemma}
\newtheorem{proposition}{Proposition}
\newtheorem{theorem}{Theorem}
\newtheorem{corollary}{Corollary}

\newtheorem{definition}{Definition}

\def\done{\hspace*{\fill} $\framebox[2mm]{}$}

\usepackage{cite}
\usepackage{amsmath}
\usepackage{graphicx}
\usepackage{array}
\usepackage{hyperref}
\usepackage{multirow}
\usepackage{color}
\begin{document}
\bibliographystyle{IEEEtran}
\title{On Capacity and Delay of Multi-channel Wireless Networks with Infrastructure Support}

\author{Hong-Ning Dai, Raymond Chi-Wing Wong, Hao Wang}
\maketitle

\begin{abstract}

In this paper, we propose a novel multi-channel network with infrastructure support, called an \textit{MC-IS} network, which has not been studied in the literature. To the best of our knowledge, we are the first to study such an \textit{MC-IS} network. Our proposed \textit{MC-IS} network has a number of advantages over three existing conventional networks, namely a single-channel wireless ad hoc network (called an \textit{SC-AH} network), a multi-channel wireless ad hoc network (called an \textit{MC-AH} network) and a single-channel network with infrastructure support (called an \textit{SC-IS} network). In particular, the \textit{network capacity} of our proposed \textit{MC-IS} network is $\sqrt{n \log n}$ times higher than that of an \textit{SC-AH} network and an \textit{MC-AH} network and the same as that of an \textit{SC-IS} network, where $n$ is the number of nodes in the network. The \textit{average delay} of our \textit{MC-IS} network is $\sqrt{\log n/n}$ times lower than that of an \textit{SC-AH} network and an \textit{MC-AH} network, and $\min\{C_I,m\}$ times lower than the average delay of an \textit{SC-IS} network, where $C_I$ and $m$ denote the number of channels dedicated for infrastructure communications and the number of interfaces mounted at each infrastructure node, respectively. Our analysis on an \emph{MC-IS} network equipped with \emph{omni-directional antennas} only has been extended to an \emph{MC-IS} network equipped with \emph{directional antennas} only, which are named as an \emph{MC-IS-DA} network. We show that an \emph{MC-IS-DA} network has an even lower delay of $\frac{c}{\lfloor \frac{2\pi}{\theta}\rfloor \cdot C_I}$ compared with an \emph{SC-IS} network and our \emph{MC-IS} network. For example, when $C_I=12$ and $\theta=\frac{\pi}{12}$, an \emph{MC-IS-DA} can further reduce the delay by 24 times lower that of an \emph{MC-IS} network and reduce the delay by 288 times lower than that of an \emph{SC-IS} network.

\end{abstract}


\IEEEpeerreviewmaketitle

\section{Introduction}
\label{sec:intro}

How to improve the network performance, in terms of the network capacity and the average delay, has been a key issue in recent studies \cite{NLu:TCST14}. Conventional wireless networks typically consist of nodes that share one single channel for communications. It is found in \cite{Gupta:Kumar,gamal:2004,gamal:TIT2006} that in a random\footnote{There are two kinds of network placements: (a) a \textit{random network}, in which $n$ nodes are randomly placed, and the destination of a flow is also randomly chosen and (b) an \textit{arbitrary network}, in which the location of nodes, and traffic patterns can be optimally controlled. We only consider the random network in this paper.} ad hoc network with $n$ nodes, each node has a throughput capacity of $\Theta(W/\sqrt{n\log n})$ (where $W$ is the total network bandwidth) and the average delay of this network is $\Theta(\sqrt{n/\log n})$. When the number of nodes increases, the per-node throughput decreases and the average delay increases. One major reason is that all the nodes within the network share the \textit{same} medium. When a node transmits, its neighboring nodes in the same channel are prohibited from transmitting to avoid interference. Besides, multi-hop and short-ranged communications are preferred in this network in order to minimize the interference and achieve the high network capacity \cite{Gupta:Kumar}. However, the multi-hop communications inevitably lead to the high end-to-end delay. Furthermore, every node equipped with a single interface cannot transmit and receive at the same time (i.e., the half-duplex constraint). We name this single-channel ad hoc network as an \textit{SC-AH} network.

One approach to improve the network performance is to use \textit{multiple channels} instead of a single channel in a wireless network. The experimental results of \cite{Raniwala:infocom2005,So:mobihoc04,Bahl:mobicom2004,Draves:mobicom2004,Kyasanur:mobicom2005,hndai:infocom2008} show that using multiple channels can significantly improve the network throughput. One possible reason for the improvement is that using multiple channels can separate multiple concurrent transmissions in frequency domains so that the interference can be mitigated. Another reason is that multiple simultaneous transmissions/receptions are supported by \textit{multiple network interfaces} mounted at a wireless node, consequently leading to the improved frequency reuse and the increased throughput. However, it is shown in \cite{Gupta:Kumar} \cite{Kyasanur:mobicom2005} that each channel (or up to $O(\log n)$ channels) must be utilized by a dedicated interface at a node in order to fully utilize all the channels simultaneously so that the network capacity can be maximized. When the condition is not fulfilled, the capacity degrades significantly. Besides, the average delay of an \textit{MC-AH} network is also $\Theta(\sqrt{n/\log n})$, which increases significantly with the increased number of nodes. We call this multi-channel wireless ad hoc network as an \textit{MC-AH} network.

Recent studies \cite{bliu:infocom2003,Kozat:mobicom2003,Zemlianov:jsac05,bliu:mobihoc2007,panli:jsac09,XWang:TC2010} investigated the performance improvement by adding a number of infrastructure nodes to a wireless network. Specifically, as shown in \cite{bliu:infocom2003,panli:jsac09}, deploying infrastructure nodes in the wireless network can significantly improve the network capacity and reduce the average delay. But, every node (both a common node and an infrastructure node) in such a network equipped with a single interface cannot transmit and receive at the same time (i.e., the half-duplex constraint is still in place). Besides, only one single channel is used in such a network. We call this single-channel networks with infrastructure support as an \textit{SC-IS} network. 

In this paper, we propose a novel multi-channel network with infrastructure support that overcomes the above drawbacks of existing networks. This network consists of \textit{common nodes}, each of which has a single interface, and \textit{infrastructure nodes} (or base stations), each of which has multiple interfaces. Both common nodes and base stations can operate on different channels. This multi-channel wireless network with infrastructure support is called an \textit{MC-IS} network that has the following characteristics.
\begin{itemize}
	\item Each common node is equipped with a single network interface card (NIC). Each base station is equipped with multiple NICs.
	\item There are multiple non-overlapping channels available. Each NIC at either a common node or a base station can switch to different channels quickly (so we can ignore the switching delay of NICs). 
	\item Base stations are connected via a \textit{wired} network that has much higher bandwidth than a wireless network.
	\item Each common node with a single NIC can communicate with either another common node or a base station, where a communication with another common node is called an ad-hoc communication and a communication with a base station is called an infrastructure communication. But, a common node supports only one transmission or one reception at a time. Besides, it cannot simultaneously transmit and receive (i.e., it is in a \textit{half-duplex} mode). 
	\item Each base station with multiple NICs can communicate with more than one common node. In addition, a base station can also work in a \textit{full-duplex} mode, i.e., transmissions and receptions can occur in parallel. 
\end{itemize}

\begin{table}[t!]
\centering
\renewcommand{\arraystretch}{1.1}
\caption{Comparison with other existing wireless networks}
\label{tab:networks}
\begin{tabular}{|c|c|c|}
\hline
 &\textit{Pure Ad Hoc} & \textit{Ad Hoc with Infrastructure} \\
\hline
\hline
 \textit{Single Channel} & \textit{SC-AH} networks  & \textit{SC-IS} networks \\
& \cite{Gupta:Kumar,gamal:2004,gamal:TIT2006} & \cite{bliu:infocom2003,panli:jsac09,Kozat:mobicom2003,Zemlianov:jsac05,XWang:TC2010,bliu:mobihoc2007,panli:infocom10,Xie:ICC12} \\ 
\hline
 \textit{Multiple Channels} & \textit{MC-AH} networks & \textit{MC-IS} networks \\
 &  \cite{Raniwala:infocom2005,So:mobihoc04,Bahl:mobicom2004,Draves:mobicom2004,Kyasanur:mobicom2005,hndai:infocom2008} & (this paper)\\ 
 \hline
\end{tabular}
\end{table}

In fact, our proposed \emph{MC-IS} networks have provided a solution to the new applications, such as \emph{Device-to-Device} (D2D) networks \cite{Asadi:CST14}, wireless sensor networks (WSNs), smart grid, smart home and e-health systems \cite{YZhang:IEEENet12,YYan:TCST13}. For example, the theoretical analysis on the throughput and the delay of our \emph{MC-IS} networks can be used to analyze the performance of the \emph{overlaid} D2D networks (details can be found in Section \ref{subsec:implications}). 

Table \ref{tab:networks} compares our proposed \textit{MC-IS} networks with other existing networks, where one can observe that \textit{MC-IS} networks can fully exploit the benefits of both \textit{MC-AH} networks and \textit{SC-IS} networks and can potentially have a better network performance (in terms of the network capacity and the delay) than other existing networks. However, to the best of our knowledge, \textit{there is no theoretical analysis on the capacity and the average delay of an \textit{MC-IS} network}. The goal of this paper is to investigate the performance of an \textit{MC-IS} network and to explore the advantages of this network. The primary research contributions of our paper are summarized as follows.
\begin{enumerate}
	\item[(1)] We formally identify an \textit{MC-IS} network that characterizes the features of \textit{multi-channel} wireless networks with \textit{infrastructure support}. {\it To the best of our knowledge, the capacity and the average delay of an \textit{MC-IS} network have not been studied before}.
	\item[(2)] We propose a \textit{general} theoretical framework to analyze the capacity and the average delay. We show that other existing networks can be regarded as special cases of our \textit{MC-IS} network in our theoretical framework. Besides, we find that our \emph{MC-IS} networks are limited by \emph{four requirements} (to be defined in Section \ref{sec:main}) \emph{simultaneously} but the existing networks are only limited by subsets of them (not all of them). This means that studying the performance of our \textit{MC-IS} networks is more challenging but it is more useful and realistic to consider four requirements simultaneously since they exist naturally in real life applications. 
	\item[(3)] Our proposed \textit{MC-IS} network has a lot of advantages over existing related networks. In particular, an \textit{MC-IS} network can achieve the \textit{optimal} per-node throughput $W$, which is $\sqrt{n \log n}$ times higher than that of an \textit{SC-AH} network and an \textit{MC-AH} network and the same as that of an \textit{SC-IS} network, while maintaining the smallest delay, which is $\sqrt{\log n/n}$ times lower than that of an \textit{SC-AH} network and an \textit{MC-AH} network, and $\min\{C_I,m\}$ times lower than that of an \textit{SC-IS} network. The performance improvement mainly owes to the multiple interfaces at a base station, compared with a single interface at a base station in \textit{SC-IS} networks. As a result, our \textit{MC-IS} networks have a better performance than \textit{SC-IS} networks though the theoretical analysis is also more complicated than that of \textit{SC-IS} networks.
	\item[(4)] We also extend our \emph{MC-IS} networks with the consideration of using \emph{directional antennas} instead of \emph{omni-directional antennas}. Specifically, all aforementioned networks (i.e., \emph{SC-AH} networks, \emph{MC-AH} networks, \emph{SC-IS} networks and our \emph{MC-IS} networks) are equipped with omni-directional antennas but the extended \emph{MC-IS} networks have both the base stations and all common nodes equipped with \emph{directional} antennas. We name the extended \emph{MC-IS} networks as \emph{MC-IS-DA} networks. We show that an \emph{MC-IS-DA} network can have an even lower delay of $\frac{c}{\lfloor \frac{2\pi}{\theta}\rfloor \cdot C_I}$ compared with both an \emph{MC-IS} network and an \textit{SC-IS} network, where $\theta$ is the beamwidth of a directional antenna mounted at the base station (usually $\theta < 2 \pi$). Consider the case of $C_I=12$ and $\theta=\frac{\pi}{12}$ that is feasible in Millimeter-Wave systems \cite{Rappaport:icc15}. An \emph{MC-IS-DA} can further reduce the delay by 24 times lower than that of an \emph{MC-IS} network and reduce the delay by 288 times lower than that of an \emph{SC-IS} network.
\end{enumerate}

The remainder of the paper is organized as follows. Section \ref{sec:related} presents a survey on the related studies to our \textit{MC-IS} network. We present the models used in this paper in Section \ref{sec:models}. Section \ref{sec:main} then summarizes our main results. We next derive the capacity and the delay contributed by \textit{ad hoc communications} in an \textit{MC-IS} network in Section \ref{sec:ad-hoc}. Section \ref{sec:infra} presents the capacity and the delay contributed by \textit{infrastructure communications} in an \textit{MC-IS} network. We extend our analysis with the consideration of directional antennas as well as the mobility and provide the implications of our results in Section \ref{sec:discussion}. Finally, we conclude the paper in Section \ref{sec:conclusion}.

\section{Related Works}
\label{sec:related}

We summarize the related works to our study in this section. The first network related to our proposed \textit{MC-IS} network is an \textit{SC-AH} network. An \textit{SC-AH} network has a poor performance due to the following reasons: (i) the interference among multiple concurrent transmissions, (ii) the number of simultaneous transmissions on a single interface and (iii) the multi-hop communications \cite{Gupta:Kumar,gamal:2004,gamal:TIT2006}.

The second network related to our \textit{MC-IS} network is an \textit{MC-AH} network, in which multiple channels instead of a single channel are used. Besides, each node in such a network is equipped with multiple network interfaces instead of single network interface. This network has a higher throughput than an \textit{SC-AH} network because each node can support multiple concurrent transmissions over multiple channels. However, this network suffers from the high delay and the increased deployment complexity. The average delay of an \textit{MC-AH} network is the same as that of an \textit{SC-AH} network, which increases significantly with the number of nodes. The deployment complexity is mainly due to the condition \cite{Kyasanur:mobicom2005} that each channel (up to $O(\log n)$ channels) must be utilized by a dedicated interface at a node so that all the channels are fully utilized simultaneously and thus the network capacity can be maximized. When the condition is not fulfilled, the capacity degrades significantly.

The third network related to our \textit{MC-IS} network is an \textit{SC-IS} network \cite{bliu:infocom2003,Kozat:mobicom2003,Zemlianov:jsac05,bliu:mobihoc2007,panli:jsac09,XWang:TC2010,panli:infocom10,Devu:INFOCOM2011,Xie:ICC12}. It is shown in \cite{bliu:infocom2003,panli:jsac09} that an \textit{SC-IS} network can significantly improve the network capacity and reduce the average delay. However, an infrastructure node in such a network equipped with a single interface cannot transmit and receive at the same time (i.e., the half-duplex constraint is still enforced). Thus, the communication delay in such an \textit{SC-IS} network is still not minimized. Besides, such \textit{SC-IS} networks also suffer from the poor spectrum reuse.

The fourth network related to our \textit{MC-IS} network is a multi-channel wireless mesh network with infrastructure support (called an \textit{MC-Mesh-IS} network) \cite{PZhou:TMC08,Mansoori:WPC13,WFu:TMC13,Mansoori:performance13,Chieochan:TMC2013,Mansoori:TMC2015}, which is the evolution of multi-channel multi-interface wireless mesh networks (called an \textit{MC-Mesh} network) \cite{Akyildiz:2005,Kodialam:2005}. An \textit{MC-Mesh-IS} network is different from our \textit{MC-IS} network due to the following characteristics of an \textit{MC-Mesh-IS} network: 
\begin{itemize}
\item [(i)] a typical \textit{MC-Mesh-IS} network consists of {\it mesh clients}, {\it mesh routers} and {\it mesh gateways} while an \textit{MC-IS} network consists of common nodes and infrastructure nodes.
\item [(ii)] different types of communications exist in the multi-tier hierarchical \textit{MC-Mesh-IS} network, which are far more complicated than an \textit{MC-IS} network. For example, there are communications between mesh clients, communications between mesh gateways, and communications between a mesh gateway and a mesh router.
\item [(iii)] an \textit{MC-Mesh-IS} network uses wireless links to connect the {\it backbone} networks (corresponding to the infrastructure network in an \textit{MC-IS} network). As a result, the assumption of the unlimited capacity and the interference-free infrastructure communications in an \textit{MC-IS} network does not hold for an \textit{MC-Mesh-IS} network.
\item [(iv)] the traffic source of an \textit{MC-Mesh-IS} network is either from a mesh client or from the Internet while the traffic always originates from the same network in an \textit{MC-IS} network.
\end{itemize}

Therefore, the analytic framework on the capacity and the delay of such \textit{MC-Mesh-IS} networks is significantly different from that of an \textit{MC-IS} network. An interesting question is whether using directional antennas in \textit{MC-Mesh-IS} networks can bring the performance improvement, which might be one of the future works.

In this paper, we analyze the capacity and the delay of an \textit{MC-IS} network. Although parts of the results on the analysis on the capacity and the delay contributed by ad hoc communications have appeared in \cite{hndai:MC-IS}, our analysis in this paper significantly differs from the previous work in the following aspects:
\begin{itemize}
\item We derive the capacity and the delay of an \textit{MC-IS} network contributed by infrastructure communications in this paper while \cite{hndai:MC-IS} only addresses the capacity and the delay contributed by ad hoc communications.
\item We fully investigate the capacity and the delay of an \textit{MC-IS} network with consideration of both infrastructure communications and ad hoc communications. Specifically, we also analyze the average delay and the optimality of our results, all of which have not been addressed in \cite{hndai:MC-IS}.
\item We also compare our results with other existing networks, such as an \textit{SC-AH} network, an \textit{MC-AH} network and an \textit{SC-IS} network and analyze the generality of our \textit{MC-IS} network in this paper.
\item We extend our analysis with consideration of using directional antennas in an \textit{MC-IS} network. Discussions on the mobility are also presented in this paper (see Section \ref{sec:discussion} for more details).
\end{itemize}

\section{Models}
\label{sec:models}

We adopt the asymptotic notations \cite{Cormen:2009} in this paper. We then describe the \textit{MC-IS} network model in Section \ref{subsec:MC-IS}. Section \ref{subsec:definitions} next gives the definitions of the throughput capacity and the delay. 

\subsection{MC-IS Network Model}
\label{subsec:MC-IS}

\begin{figure}[t]
\centering
\includegraphics[width=8.0cm]{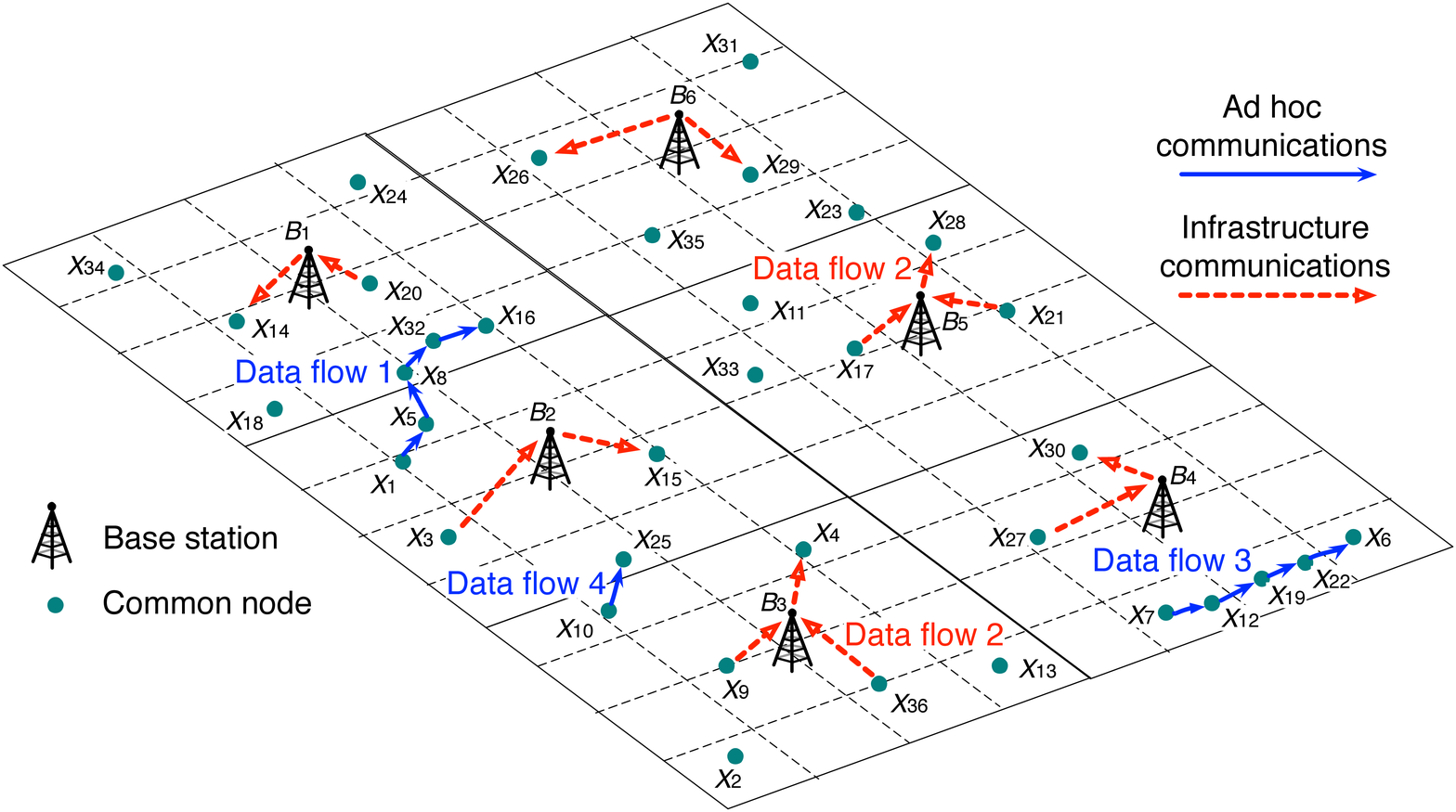}
\caption{Network topology of an \textit{MC-IS} network}
\label{fig:network}
\end{figure}

Take Fig. \ref{fig:network} as an example of \textit{MC-IS} networks. In this network, $n$ common nodes are randomly, uniformly and independently distributed on a unit square plane $A$. Each node is mounted with a single interface that can switch to one of $C$ available channels. Each node can be a data source or a destination. All the nodes are homogeneous, which means that they have the same transmission range. In addition, there are $b$ infrastructure nodes, which are also called \textit{base stations} interchangeably throughout the whole paper. We assume that $b$ can be expressed as a square of a constant $b_0$ (i.e., $b_0^2$) where $b_0$ is an integer in order to simplify our discussion. Each base station is equipped with $m$ interfaces and each interface is associated with a single omni-directional antenna, which can operate on one of $C$ channels. The plane $A$ is evenly partitioned into $b$ equal-sized squares, which are called \textit{BS-cells}. Similar to \cite{bliu:infocom2003,panli:jsac09,XWang:TC2010}, we also assume that a base station is placed at the center of each \textit{BS-cell}. Unlike a node, a base station is neither a data source nor a destination and it only helps forwarding data for nodes. All the base stations are connected through a wired network without \textit{capacity constraint} and \textit{delay constraint}.

There are two kinds of communications in an \textit{MC-IS} network: (i) \textit{Ad hoc communications} between two nodes, which often proceed in a multi-hop manner; (ii) \textit{Infrastructure communications} between a node and a base station, which span a single hop. An infrastructure communication consists of an \textit{uplink} infrastructure communication from a node to a base station, and a \textit{downlink} infrastructure communication from a base station to a node.

In the following, we describe two major components for network communications. The first component is the routing strategy. The second component is the interference model.

\subsubsection{Routing Strategy}
In this paper, we consider the \textit{$H$-max-hop routing strategy}, in which, if the destination is located within $H$ ($H\geq 1$) hops from the source node, data packets are transmitted through ad hoc communications. Otherwise, data packets are forwarded to the base station through infrastructure communications (i.e., the uplink infrastructure communication). The base station then relays the packets through the wired network. After the packets arrive at the base station that is closest to the destination node, the base station then forwards the packets to the destination node (i.e., the downlink infrastructure communication). Take Fig. \ref{fig:network} as the example again. Data flow 1 starts from node $X_1$ to node $X_{16}$ in the multi-hop ad hoc manner since node $X_{16}$ is within $H$ hops from node $X_1$. With regard to Data flow 2, since destination node $X_{28}$ is far from source node $X_{36}$, data packets are transmitted from source node $X_{36}$ to its nearest base station $B_3$ first and then are forwarded through the wired network till reaching base station $B_5$ that finally sends the data packets to destination node $X_{28}$. 

The \textit{$H$-max-hop routing strategy} can avoid the problem that arises by using the \textit{$k$-nearest-cell routing strategy} in the case of two nodes near the boundary of two adjacent \emph{BS-cells}. For example, Data flow 4 as shown in Fig. \ref{fig:network} starting from node $X_{10}$ to destination node $X_{25}$ will be transmitted in one hop by ad hoc communications according our \textit{$H$-max-hop routing strategy}. However, in the \textit{$k$-nearest-cell routing strategy} \cite{bliu:infocom2003}, node $X_{10}$ has to transmit to its nearest \emph{BS} (i.e., $B_3$) first and then $B_3$ forwards the data packets through the wired network till they reach $B_2$, which is the nearest \emph{BS} to node $X_{25}$. This problem may result in inefficient use of bandwidth resources. 

It is obvious that \textit{when there is an uplink communication, there is always a downlink communication}. We then divide the total bandwidth of $W$ bits/sec into three parts: (1) $W_A$ for ad hoc communications, (2) $W_{I,U}$ for uplink infrastructure communications and (3) $W_{I,D}$ for downlink infrastructure communications. Since $W_{I,U}$ is equal to $W_{I,D}$, it is obvious that $W=W_A+W_{I,U}+W_{I,D}=W_A+2W_{I,U}$. To simplify our analysis, we use $W_I$ to denote either $W_{I,U}$ or $W_{I,D}$. Corresponding to the partition of the bandwidth, we also split the $C$ channels into two disjoint groups $C_A$ and $C_I$, in which $C_A$ channels are dedicated for ad hoc communications and $C_I$ channels are dedicated for infrastructure communications. Thus, $C=C_A+C_I$. Besides, each base station is mounted with $m$ NICs, which serve for both the uplink traffic and the downlink traffic. It is obvious that the number of NICs serving for the uplink traffic is equal to the number of NICs serving for the downlink traffic. So, $m$ must be an even number.

\subsubsection{Interference model}
In this paper, we consider the \textit{interference} model \cite{Gupta:Kumar,bliu:infocom2003,Kozat:mobicom2003,Kyasanur:mobicom2005,Zemlianov:jsac05,panli:jsac09}. When node $X_i$ transmits to node $X_j$ over a particular channel, the transmission is successfully completed by node $X_j$ if no node within the transmission range of $X_j$ transmits over the same channel. Therefore, for any other node $X_k$ simultaneously transmitting over the same channel, and any guard zone $\Delta>0$, the following condition holds.
\begin{displaymath}
\textrm{dist}(X_k,X_j)\geq(1+\Delta)\textrm{dist}(X_i,X_j)
\end{displaymath}
where $\textrm{dist}(X_i,X_j)$ denotes the distance between two nodes $X_i$ and $X_j$. Note that the \textit{physical interference} model \cite{Gupta:Kumar} is ignored in this paper since the physical model is equivalent to the interference model when the \textit{path loss exponent} is greater than two (it is common in a real world \cite{Gupta:Kumar,Rappaport:2002}).

The interference model applies for both ad hoc communications and infrastructure communications. Since ad hoc communications and infrastructure communications are separated by different channels (i.e., $C_A$ and $C_I$ do not overlap each other), the interference only occurs either between two ad hoc communications or between two infrastructure communications. 

\subsection{Definitions of Throughput Capacity and Delay}
\label{subsec:definitions}
The notation of throughput of a transmission from a node $X_i$ to its destination node $X_j$ is usually defined as the number of bits that can be delivered from $X_i$ to $X_j$ per second. The \textit{aggregate throughput capacity} of a network is defined to be the total throughput of all transmissions in the network. The \textit{per-node throughput capacity} of a network is defined to be its aggregate throughput capacity divided by the total number of transmissions (or all nodes involved in transmissions). In this paper, we mainly concentrate on the \textit{per-node throughput capacity} and the \textit{average delay}, which are defined as follows. 

\begin{definition}
\textit{Feasible per-node throughput}. For an \textit{MC-IS} network, a throughput of $\lambda$ (in bits/sec) is \textit{feasible} if by ad hoc communications or infrastructure communications, there exists a \textit{spatial and temporal scheme}, within which each node can send or receive $\lambda$ bits/sec on average. 
\end{definition} 

\begin{definition}
\textit{Per-node throughput capacity of an \textit{MC-IS} network} with the throughput of $\lambda$ is of order $\Theta(g(n))$ bits/sec if there are deterministic constants $h>0$ and $h'<+\infty$ such that
\begin{displaymath}
\begin{array}{ll}
\lim_{n \rightarrow \infty} & P(\lambda= h g(n) \textrm{ is feasible})=1 \textrm{ and} \nonumber\\
\lim_{n \rightarrow \infty} \inf &  P(\lambda= h' g(n) \textrm{ is feasible})<1.
\end{array}
\end{displaymath}
\end{definition} 

In this paper, the per-node throughput capacity of an \textit{MC-IS} network is expressed by $\lambda = \lambda_a+\lambda_i$, where $\lambda_a$ and $\lambda_i$ denote the throughput capacity contributed by the ad hoc communications and the infrastructure communications, respectively. Besides, we use $T$, $T_A$, $T_I$ to denote the \textit{feasible aggregate throughput}, the \textit{feasible} aggregate throughput contributed by \textit{ad hoc} communications, and the \textit{feasible} aggregate throughput contributed by \textit{infrastructure} communications, respectively.

\begin{definition}
\textit{Average Delay of an \textit{MC-IS} network}. The \textit{delay of a packet} is defined as the time that it takes for the packet to reach its destination after it leaves the source \cite{gamal:TIT2006}. After averaging the delay of all the packets transmitted in the whole network, we obtain the \textit{average delay} of an \textit{MC-IS} network, denoted by $D$.
\end{definition}

The average delay of an \emph{MC-IS} network is expressed by $D=D_a+D_i$, where $D_a$ and $D_i$ denote the delay contributed by ad hoc communications and the delay contributed by infrastructure communications, respectively. To derive the average delay in this paper, we consider the {\it fluid model} proposed by A. El. Gamal et al. in \cite{gamal:2004,gamal:TIT2006}. In this model, the packet size is allowed to be arbitrarily small so that the time taken for transmitting a packet may only occupy a small fraction of one time slot, implying that multiple packets can be transmitted within one time slot. The fluid model can be easily extended to the case of the packet with constant size as shown in \cite{gamal:TIT2006II}. Note that we do not count the delay caused by the infrastructure communications within the wired network. Besides, we also ignore the queuing delay in this model.

In order to compare the optimality of our results with the existing ones, we introduce the \emph{optimal} per-node throughput capacity $\lambda_{opt}$, which is the maximum achievable per-node throughput capacity, and the \emph{optimal} average delay $D_{opt}$, which is the average delay when the optimal per-node throughput capacity $\lambda_{opt}$ is achieved. 

\section{main results}
\label{sec:main}
We first present the four requirements that limit the capacity of an \textit{MC-IS} network in Section \ref{subsec:fourR}. Section \ref{subsec:summary} then gives the main results.

\subsection{Four Requirements}
\label{subsec:fourR}

\begin{figure*}[t]
\centering
\includegraphics[width=11cm]{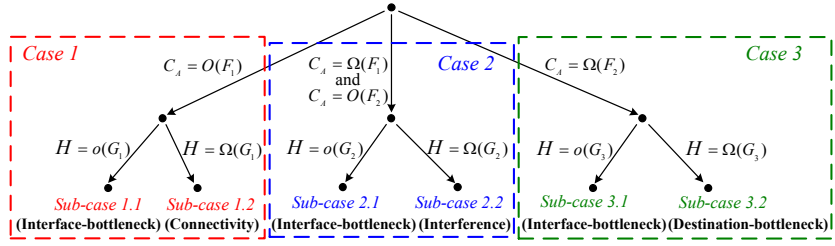}
\caption{All possible sub-cases considered}
\label{fig:cases}
\end{figure*}

We have found that the capacity of an \textit{MC-IS} network is mainly limited by \emph{four requirements simultaneously}: (i) \textit{Connectivity requirement} - the need to ensure that the network is connected so that each source node can successfully communicate with its destination node; (ii) \textit{Interference requirement} - two receivers simultaneously receiving packets from two different transmitters must be separated with a minimum distance to avoid the interference between the two transmissions for the two receivers; (iii) \textit{Destination-bottleneck requirement} - the maximum amount of data that can be simultaneously received by a destination node; (iv) \textit{Interface-bottleneck requirement} - the maximum amount of data that an interface can simultaneously transmit or receive. Besides, each of the four requirements dominates the other three requirements in terms of the throughput of the network under different conditions on $C_A$ and $H$. 

\emph{Our findings are significantly different from the previous studies in {\it SC-AH} networks, {\it MC-AH} networks and {\it SC-IS} networks, which are limited by only subsets of the four requirements} (but only some of them and not all). For example, the capacity of {\it SC-AH} networks and {\it SC-IS} networks is limited by \textit{Connectivity requirement} and \textit{Interference requirement} as shown in \cite{Gupta:Kumar} and \cite{bliu:infocom2003} while the capacity of {\it MC-AH} networks is limited by \textit{Connectivity requirement}, \textit{Interference requirement} and \textit{Interface-bottleneck requirement} \cite{Kyasanur:mobicom2005} (under random network placement). As a result, our analysis on an \textit{MC-IS} network is far more \emph{challenging} than those in the previous studies.

More specifically, as shown in Fig. \ref{fig:cases}, $C_A$ can be partitioned into 3 cases: \emph{Case 1} corresponding to the case when $C_A=O(F_1)$, \emph{Case 2} corresponding to the case when $C_A=\Omega(F_1)$ and $C_A=O(F_2)$, and \emph{Case 3} corresponding to the case when $C_A=\Omega(F_2)$, where $F_1=\log n$ and $F_2=n(\frac{\log{\log{(H^2 \log n)}}}{\log{(H^2 \log n)}})^2$.

Under each of the above cases, $H$ can be partitioned into two sub-cases. Under \emph{Case 1}, $H$ is partitioned into 2 sub-cases, namely \emph{Sub-case 1.1} and \emph{Sub-case 1.2}. \emph{Sub-case 1.1} is when $H=o(G_1)$ and \emph{Sub-case 1.2} is when $H=\Omega(G_1)$, where $G_1=n^{\frac{1}{3}}/\log^{\frac{2}{3}} n$. Under \emph{Case 2}, $H$ is partitioned into 2 sub-cases, namely \emph{Sub-case 2.1} and \emph{Sub-case 2.2}. \emph{Sub-case 2.1} is when $H=o(G_2)$ and \emph{Sub-case 2.2} is when $H=\Omega(G_2)$, where $G_2=n^{\frac{1}{3}}C^{\frac{1}{6}}_A/\log^{\frac{1}{2}} n$. Under \emph{Case 3}, $H$ is partitioned into 2 sub-cases, namely \emph{Sub-case 3.1} and \emph{Sub-case 3.2}. \emph{Sub-case 3.1} is when $H=o(G_3)$ and \emph{Sub-case 3.2} is when $H=\Omega(G_3)$, where $G_3=n^{\frac{1}{2}}/\log^{\frac{1}{2}} n$. Fig. \ref{fig:cases} shows all possible sub-cases we consider.

Each requirement dominates the other at least one sub-case under different conditions as follows.
\begin{itemize}
\item \textit{Connectivity Condition}: corresponding to \emph{Sub-case 1.2} in which \emph{Connectivity requirement} dominates.
\item \textit{Interference Condition}: corresponding to \emph{Sub-case 2.2} in which \emph{Interference requirement} dominates.
\item \textit{Destination-bottleneck Condition}: corresponding to \emph{Sub-case 3.2} in which \emph{Destination-bottleneck requirement} dominates.
\item \textit{Interface-bottleneck Condition}: corresponding to \emph{Sub-case 1.1}, \emph{Sub-case 2.1}, or \emph{Sub-case 3.1}, in which \emph{Interface-bottleneck requirement} dominates.
\end{itemize}

\subsection{Summary of Results}
\label{subsec:summary}

We summarize the main results as follows.

\emph{1. Throughput and Delay for an MC-IS network}

\begin{theorem}
\label{th:throughput}
The per-node throughput $\lambda$ for an \textit{MC-IS} network has four regions as follows.
\begin{enumerate}
\item [i)] When \textit{Connectivity Condition} is satisfied, $\lambda = \Theta\big(\frac{W_A}{ H \log n}\big)+\Theta\big(\min\{\frac{b}{n},\frac{bm}{nC_I}\}W_I\big)$, where $\lambda_a = \Theta\big(\frac{W_A}{ H \log n}\big)$ and $\lambda_i = \Theta\big(\min\{\frac{b}{n},\frac{bm}{nC_I}\}W_I\big)$;
\item [ii)] When \textit{Interference Condition} is satisfied, $\lambda = \Theta\Big(\frac{ W_A}{C_A^{\frac{1}{2}} H \log^{\frac{1}{2}}n}\Big)+\Theta(\min\{\frac{b}{n},\frac{bm}{nC_I}\}W_I)$, where $\lambda_a =\Theta\Big(\frac{ W_A}{C_A^{\frac{1}{2}} H \log^{\frac{1}{2}}n}\Big)$ and $\lambda_i = \Theta\big(\min\{\frac{b}{n},\frac{bm}{nC_I}\}W_I\big)$;
\item [iii)] When \textit{Destination-bottleneck Condition} is satisfied, $\lambda = \Theta\bigg(\frac{n^{\frac{1}{2}} \log \log(H^2 \log n)W_A}{C_A H \log^{\frac{1}{2}}n \cdot \log(H^2 \log n)}\bigg)+\Theta(\min\{\frac{b}{n},\frac{bm}{nC_I}\}W_I)$, where $\lambda_a = \Theta\bigg(\frac{n^{\frac{1}{2}} \log \log(H^2 \log n)W_A}{C_A H \log^{\frac{1}{2}}n \cdot \log(H^2 \log n)}\bigg)$ and $\lambda_i = \Theta\big(\min\{\frac{b}{n},\frac{bm}{nC_I}\}W_I\big)$;
\item [iv)] When \textit{Interface-bottleneck Condition} is satisfied, $\lambda = \Theta\bigg(H^2 \frac{\log n}{n} \cdot \frac{W_A}{C_A}\bigg)+\Theta(\min\{\frac{b}{n},\frac{bm}{nC_I}\}W_I)$, where $\lambda_a = \Theta\bigg(H^2 \frac{\log n}{n} \cdot \frac{W_A}{C_A}\bigg)$ and $\lambda_i = \Theta\big(\min\{\frac{b}{n},\frac{bm}{nC_I}\}W_I\big)$.
\end{enumerate}
\end{theorem}

\begin{theorem}
\label{th:delay}
The average delay of all packets in an \textit{MC-IS} network is $D=\Theta\Big(\frac{H^3 \log n}{n}\Big) + \Theta\Big(\frac{c}{\min\{C_I,m\}}\Big)$, where $D_a=\Theta\Big(\frac{H^3 \log n}{n}\Big)$ and $D_i=\Theta\Big(\frac{c}{\min\{C_I,m\}}\Big)$.
\end{theorem}

\emph{2. Overview of Our Proof}

Since ad hoc communications and infrastructure communications are carried in two disjoint channel groups $C_A$ and $C_I$, we will derive the bounds on the capacity and the delay contributed by the two communications separately. In particular, we first obtain the bounds on the the capacity contributed by ad hoc communications in Section \ref{sec:ad-hoc}. More specifically, we will derive the upper bounds on the capacity by consideration of the aforementioned four requirements and then prove the lower bounds by constructing the cells, designing routing scheme and TDMA scheme properly. Although our approach is the integration of the previous studies on \textit{SC-IS} networks \cite{panli:jsac09} and \textit{MC-AH} networks \cite{Kyasanur:mobicom2005}, our solution is non-trivial due to the following reasons: (i) the capacity of \textit{MC-IS} networks is limited by the aforementioned four conditions simultaneously while those of \textit{SC-IS} networks and \textit{MC-AH} networks are only limited by subsets of the four conditions; (ii) as a result, we need to redesign the cell construction, the routing scheme and the scheduling scheme based on various factors (such as $H$, $C_A$ and $n$), which are not straight-forward. Details about our proof on ad hoc communications will be given in Section \ref{sec:ad-hoc}. We will next derive the capacity contributed by infrastructure communications in Section \ref{sec:infra}. Similarly, we need to construct \textit{BS-cells}, design routing scheme and TDMA scheme in this phrase while these constructions are different from those of ad hoc communications. The complete proof of Theorem \ref{th:throughput} and Theorem \ref{th:delay} will be given in Section \ref{sec:infra}.

\emph{3. Generality of \textit{MC-IS} Networks}

Our proposed \textit{MC-IS} network offers a more general theoretical framework than other existing networks. In particular, other networks such as an \textit{SC-AH} network \cite{Gupta:Kumar}, an \textit{MC-AH} network \cite{Kyasanur:mobicom2005}, and an \textit{SC-IS} network \cite{panli:jsac09} can be regarded as special cases of our \textit{MC-IS} network under the following scenarios. 

\textit{(A) An \textit{SC-AH} network is a special case of our \textit{MC-IS} network:}
The theoretical bounds in the \textit{SC-AH} network \cite{Gupta:Kumar} are consistent with our bounds when our configuration is set to the one for the \textit{SC-AH} network. Specifically, the configuration is that $H$ is set to $\Theta(\sqrt{n/\log n})$, $C_A=1$, $W_A=W$ and $W_I=0$. In that configuration, the total bandwidth is assigned for ad hoc communications ($W_A=W$ and $W_I=0$), there is a single channel available ($C_A=1$) corresponding to that of an \textit{SC-AH} network \cite{Gupta:Kumar}. 

\textit{(B) An \textit{MC-AH} network is a special case of our \textit{MC-IS} network:}
The theoretical bounds in the \textit{MC-AH} network \cite{Kyasanur:mobicom2005} are consistent with our bounds shown in Theorem \ref{th:throughput}, when our configuration is set to the one for the \textit{MC-AH} network, in which $H$ is set to $\Theta(\sqrt{n/\log n})$, corresponding to that of an \textit{MC-AH} network \cite{Kyasanur:mobicom2005}. 

In particular, we have the following cases:
\begin{itemize}
\item Case I: when $C_A=O(\log n)$ and $H=\Theta(\sqrt{n/\log n})$ (Connectivity Condition is  satisfied), the per-node throughput $\lambda=\Theta(W/\sqrt{n \log n})$ and the average delay $D=\Theta(\sqrt{n/\log n})$, which matches the result of an \textit{MC-AH} network \cite{Kyasanur:mobicom2005};

\item Case II: when $C_A=\Omega(\log n)$ and $C_A=O\Big(n\Big(\frac{\log{\log{(H^2 \log n)}}}{\log{(H^2 \log n)}}\Big)^2\Big)$, and  $H=\Theta(\sqrt{n/\log n})$ (Interference Condition is satisfied), the per-node throughput $\lambda=\Theta(W/\sqrt{C_A n })$ and the average delay $D=\Theta(\sqrt{n/\log n})$, which matches the result of an \textit{MC-AH} network \cite{Kyasanur:mobicom2005};

\item Case III: when $C_A=\Omega\Big(n\Big(\frac{\log{\log{(H^2 \log n)}}}{\log{(H^2 \log n)}}\Big)^2\Big)$ and  $H=\Theta(\sqrt{n/\log n})$ (Destination-bottleneck Condition is satisfied), the per-node throughput $\lambda=\Theta(\frac{n \log\log n W}{C_A \log n })$ and the average delay $D=\Theta(\sqrt{n/\log n})$, which matches the result of an \textit{MC-AH} network \cite{Kyasanur:mobicom2005}.
\end{itemize}
Note that we do not consider the capacity contributed by infrastructure communications in the above four cases.

\textit{(C) An \textit{SC-IS} network is a special case of our \textit{MC-IS} network:}
Similarly, the theoretical bounds in the \textit{SC-IS} network \cite{panli:jsac09} are consistent with our bounds when our configuration is set to the one for the \textit{SC-IS} network. 

In particular, we have the following cases:
\begin{itemize}
\item Case I: when $C_A=1$ and $H=\Omega(n^{\frac{1}{3}}/\log^{\frac{2}{3}} n)$ (Connectivity Condition is satisfied), $\lambda=\Theta(\frac{W_a}{H\log n}+\frac{b}{n}W_i)$ and $D=\Theta(\frac{H^3 \log n}{n}+c)$, which matches the result of an \textit{SC-IS} network \cite{panli:jsac09};
\item Case II: when $C_A=1$ and $H=o(n^{\frac{1}{3}}/\log^{\frac{2}{3}} n)$ (Interface-bottleneck Condition is satisfied), $\lambda=\Theta(H^2 \frac{\log n}{n} \cdot \frac{W_a}{C_a}+\min\{\frac{b}{n},\frac{bm}{nC_i}\}W_I)$ and $D=\Theta(\frac{H^3 \log n}{n}+c)$, which matches the result of an \textit{SC-IS} network \cite{panli:jsac09}.
\end{itemize}

\emph{4. Optimality of Results}

We analyze the optimality of the per-node throughput capacity $\lambda$ and the average delay $D$ of an \textit{MC-IS} network. Specifically, the analysis is categorized into two cases: (1) when $\lambda_a$ dominates $\lambda_i$; (2) when $\lambda_i$ dominates $\lambda_a$.

\textit{Case 1}: when $\lambda_a$ dominates $\lambda_i$ (i.e. $W_A \rightarrow W$ and $W_I/W\rightarrow 0$).

We obtain the maximum per-node throughput capacity as the following sub-cases: (i) $\lambda=\Theta\big(\frac{W}{  H \log n}\big)$ with \emph{Connectivity} condition; (ii) $\lambda=\Theta\Big(\frac{W}{C^{\frac{1}{2}} H \log^{\frac{1}{2}} n}\Big)$ with \emph{Interference} condition; (iii) $\lambda=\Theta\bigg(\frac{n^{\frac{1}{2}} \log \log(H^2 \log n)W}{C H \log^{\frac{1}{2}}n \cdot \log(H^2 \log n)}\bigg)$ with \emph{Destination-bottleneck} condition; (iv) $\lambda=\Theta\big(\frac{H^2 W \log n}{C n}\big)$ with \emph{Interface-bottleneck} condition. In all the above sub-cases, we always have the average delay $D=\Theta\Big(\frac{H^3 \log n}{n}\Big)$. The results imply that we should assign most of channel bandwidth to ad hoc communications in order to obtain the maximum capacity and the minimum delay. However, we will show as follows that the above results are not optimal compared with \textit{Case 2}.

\textit{Case 2}: when $\lambda_i$ dominates $\lambda_a$ (i.e. $W_I \rightarrow W/2$ and $W_A/W\rightarrow 0$).

In this case, the maximum per-node throughput capacity $\lambda=\Theta(\frac{b}{n}W)$ and the average delay $D=\Theta\big(\frac{c}{\min\{C_I,m\}}\big)$. It implies that when when $\lambda_i$ dominates $\lambda_a$, to maximize the capacity, most of the channel bandwidth should be assigned for infrastructure communications. At this time, increasing the number of base stations can significantly improve the network capacity. More specifically, if $b=\Omega(n)$, then $\lambda=\Theta(W)$, which is significantly higher than those in \textit{Case 1}. This is because the multi-hop ad hoc communications may lead to the capacity loss due to the higher interference of multiple ad hoc communications. Meanwhile, the minimum average delay $D$ in this case is bounded by $\Theta\big(\frac{c}{\min\{C_I,m\}}\big)$, where $c$ is a constant and $\frac{c}{\min\{C_I,m\}}$ is independent of $n$. It is obvious that $\frac{c}{\min\{C_I,m\}} = o\Big(\Theta\big(\frac{H^3 \log n}{n}\big)\Big)$ since $H$ is determined by the number of nodes $n$. Intuitively, we have much lower delay than that of \textit{Case 1}. The reason behind this lies in the higher delay brought by the multi-hop communications in \textit{Case 1}. In summary, \textit{MC-IS} networks have the \emph{optimal} per-node throughput capacity $\lambda_{opt}=\Theta(W)$ and the \emph{optimal} average delay $D_{opt}=\Theta(\frac{c}{\min\{C_I,m\}})$. 

\begin{figure}[t]
\centering
\includegraphics[width=6cm]{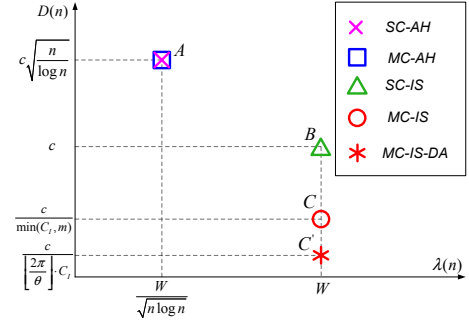}
\caption{Capacity and delay regions under different networks. The scales of the axes are in terms of the orders in $n$}
\label{fig:cap-delay}
\end{figure}

We summarize our key results and compare our results with other related networks in Fig. \ref{fig:cap-delay}. In particular, we compare an \textit{MC-IS} network with three existing networks, namely an \textit{MC-AH} network, an \textit{SC-IS} network, and an \textit{SC-AH} network, in terms of the optimal per-node throughput capacity $\lambda$ and the optimal average delay $D$ in Fig. \ref{fig:cap-delay}. As shown in Fig. \ref{fig:cap-delay}, an \textit{MC-IS} network can achieve the optimal per-node throughput capacity $\lambda_{opt}=\Theta(W)$ (point \textit{C} in Fig. \ref{fig:cap-delay}), which is $\sqrt{n \log n}$ times higher than that of an \textit{MC-AH} network and an \textit{SC-AH} network (point \textit{A} in Fig. \ref{fig:cap-delay}), and the same as that of an \textit{SC-IS} network (point \textit{B} in Fig. \ref{fig:cap-delay}). In other words, there is no capacity degradation in the optimal per-node throughput of an \textit{MC-IS} network. 

Besides, compared with other existing networks, an \textit{MC-IS} network can achieve the smallest delay $\Theta\big(\frac{c}{\min\{C_I,m\}}\big)$ (point \textit{C} in Fig. \ref{fig:cap-delay}) when the optimal per-node throughput capacity $\lambda=\Theta(W)$ is achieved. It is shown in \cite{gamal:2004,gamal:TIT2006} that in an \textit{SC-AH} network and an \textit{MC-AH} network, the increased capacity pays for the higher delay due to the multi-hop transmissions. However, an \textit{MC-IS} network and an \textit{SC-IS} network can overcome the delay penalty by transmitting packets through infrastructure, inside which there is no delay constraint. Furthermore, an \textit{MC-IS} network can achieve an even shorter delay than an \textit{SC-IS} network by using multiple interfaces at each base station, which can support multiple simultaneous transmissions. Specifically, as shown in Fig. \ref{fig:cap-delay}, an \textit{MC-IS} network (point \textit{C} in Fig. \ref{fig:cap-delay}) has a delay reduction gain of $\frac{1}{\min\{C_I,m\}}$ over an \textit{SC-IS} network (point \textit{B} in Fig. \ref{fig:cap-delay}). For example, an \textit{MC-IS} network with $C_I=m=12$ (e.g., there are $C_I=12$ non-overlapping channels in IEEE 802.11a \cite{IEEE80211a:1999}), in which we assign a dedicated interface for each channel, has a delay 12 times lower than an \textit{SC-IS} network. Besides, when we extend our analysis on an \emph{MC-IS} network equipped with \emph{omni-directional antennas} only to an \emph{MC-IS} network equipped with \emph{directional antennas} only, which are named as an \emph{MC-IS-DA} network, we can obtain an even lower delay of $\frac{c}{\lfloor \frac{2\pi}{\theta}\rfloor \cdot C_I}$ as shown in point \textit{C$'$} in Fig. \ref{fig:cap-delay}, where $\theta$ is the beamwidth of a directional antenna mounted at the base station (usually $\theta < 2 \pi$). Consider the same case of $C_I=12$ and $\theta=\frac{\pi}{12}$ that is feasible in most of mmWave systems \cite{Rappaport:icc15}. An \emph{MC-IS-DA} can further reduce the delay by 24 times lower that of an \emph{MC-IS} network and reduce the delay by 288 times lower than that of an \emph{SC-IS} network. This is because using directional antennas can concentrate the transmissions to the desired directions and can improve the spectrum reuse, potentially supporting more concurrent transmissions. Details on this extended work will be addressed in Section \ref{sec:discussion}.

\section{Capacity Contributed by Ad Hoc Communications}
\label{sec:ad-hoc}
We first derive the upper bounds on the network capacity contributed by ad hoc communications in Section \ref{sec:upper-adhoc}. Section \ref{sec:lower-adhoc} presents constructive lower bounds on the network capacity contributed by ad hoc communications, which have the same order of the upper bounds, implying that our results are quite tight. We then give the aggregate throughput capacity in Section \ref{sec:summary:ah}. Note that our derivations are significantly different from those of the existing networks, such as \emph{SC-AH} networks, \emph{MC-AH} networks and \emph{SC-IS} networks because the capacity contributed by ad hoc communications of \emph{MC-IS} networks is mainly limited by all four aforementioned requirements simultaneously (not just subsets of these requirements) and we need to establish a new theoretical framework to analyze the results. 

\subsection{Upper Bounds on Network Capacity Contributed by Ad Hoc Communications}
\label{sec:upper-adhoc}

We found that the network capacity contributed by ad hoc transmissions in an \textit{MC-IS} network, denoted by $\lambda_a$, is mainly affected by (1) \textit{Connectivity} requirement, (2) \textit{Interference} requirement, (3) \textit{Destination-bottleneck} requirement and (4) \textit{Interface-bottleneck} requirement. We first derive the upper bounds on the per-node throughput capacity under Connectivity Condition (defined in Section \ref{subsec:fourR}). Before presenting Proposition \ref{prop:regime-i}, we have Lemma \ref{lemma:H} to bound the expectation of the number of hops, which is denoted by $\overline{h}$.

\begin{lemma}
\label{lemma:H}
The expectation of the number of hops $\overline{h}$ is bounded by $\Theta(H)$.
\end{lemma}
\textbf{Proof.} We first denote $P(h=i)$ by the probability of the event that a packet traverses $h=i$ hops. According to the $H$-max-hop routing scheme, $P(h=i)$ is essentially equal to the probability that a packet traverses at most $h=i$ hops with the exclusion of the event that a packet traverses no more than $h=i-1$ hops, where $i>0$. Thus, $P(h=i)$ is equal to the ratio of the area of a disk with radius $ (i-1) \cdot r(n)$ to the area of a disk with radius $ i \cdot r(n)$, where $r(n)$ is the distance of a hop. As a result, $P(h=i)=\frac{ (i^2 - (i-1)^2) \cdot \pi r^2(n)}{\pi i^2 r^2(n)}$.

We then have
\begin{eqnarray}
\label{eqn:expofH}
\overline{h} = E(h) & = & 1 \cdot P(h=1) + 2 \cdot P(h=2) + \ldots \nonumber\\
& & + H \cdot P(h=H)\nonumber\\
& = & 1 \cdot \frac{\pi r^2(n)}{\pi H^2 r^2(n)} + 2 \cdot \frac{3 \pi r^2(n)}{\pi H^2 r^2(n)} + \ldots \nonumber\\
& & + H \cdot \frac{ (H^2 - (H-1)^2) \cdot \pi r^2(n)}{\pi H^2 r^2(n)}.
\end{eqnarray}

Since $H [H^2 - (H-1)^2]$ in Eq. (\ref{eqn:expofH}) are the series of hexagonal numbers, then Eq. (\ref{eqn:expofH}) can be simplified as follows 
\begin{eqnarray}
\label{eqn:expofH2}
\overline{h} & = & \frac{\frac{1}{6} H (H+1) (4H-1)}{H^2} = \frac{4H^3+3H^2-H}{6H^2}.
\end{eqnarray}

It is obvious that $\overline{h}$ is a function of $H$ as shown in Eq. (\ref{eqn:expofH2}). The limit of $\overline{h}(H)$ as $H$ approaches $\infty$ is 
\begin{displaymath}
\lim_{H\rightarrow \infty} \overline{h}(H) = \Theta(H),
\end{displaymath}
which can be directly derived from the definition of the asymptotic notation $\Theta(\cdot)$ and Eq. (\ref{eqn:expofH2}).
\done

We then have Proposition \ref{prop:regime-i} that bounds the per-node throughput capacity contributed by ad hoc communications under Connectivity condition:

\begin{proposition}
\label{prop:regime-i}
When Connectivity requirement dominates, the per-node throughput capacity contributed by ad hoc communications is $\lambda_a=O\big(\frac{nW_A}{H^3 \log^2 n}\big)$.
\end{proposition}
\textbf{Proof.} We first calculate the probability that a node uses the ad hoc mode to transmit, denoted by $P(AH)$, which is the probability that the destination node is located within $H$ hops away from the source node. Thus, we have 
\begin{eqnarray}
\label{eqn:prob_ah}
P(AH) = \pi H^2 r^2(n).
\end{eqnarray}

Since each source generates $\lambda_a$ bits per second and there are totally $n$ sources, the total number of bits per second served by the whole network is required to be at least $n\cdot P(AH)\cdot \overline{h} \cdot \lambda_a$. We next prove that $n\cdot P(AH)\cdot \overline{h} \cdot \lambda_a$ is bounded by $ \frac{k_1}{\Delta^2 (r(n))^2} W_A$.

We denote the maximum number of simultaneous transmissions on a particular channel by $N_{\max}$. As proved in Lemma 5.4 in \cite{Gupta:Kumar}, $N_{\max}$ is upper bounded by $\frac{k_1}{\Delta^2 (r(n))^2}$, where $k_1>0$ is a constant, independent of $n$. Note that each transmission over the $\varpi$ channel is of $W_A/C_A$ bits/sec. Adding all the transmissions taking place at the same time over all the $C_A$ channels, we have the total number of transmissions in the whole network is no more than 
\begin{displaymath}
\frac{k_1}{\Delta^2 (r(n))^2}\sum_{\varpi=1}^{C_A} \frac{W_A}{C_A} = \frac{k_1}{\Delta^2 (r(n))^2} W_A \textrm{bits/sec}.
\end{displaymath}
Therefore, we have $n\cdot P(AH)\cdot \overline{h} \cdot \lambda_a \leq \frac{k_1}{\Delta^2 (r(n))^2} W_A$.

Combining the above results with Lemma \ref{lemma:H} yields $\lambda_a \leq \frac{k_1}{\Delta^2 r^2(n)} \cdot \frac{W_A}{n \pi H^3 r^2(n) } \leq \frac{k_2 W_A}{n H^3 r^2(n)}$, where $k_2$ is a constant. 

Besides, to guarantee that the network is connected with high probability (\textit{w.h.p.})\footnote{We say that an event $e$ happens with a high probability if $P(e)\rightarrow 1$ when $n \rightarrow \infty$.}, we require $r(n)>\sqrt{\frac{\log n}{\pi n}}$ \cite{Gupta:Kumar}. Thus, we have $\lambda_a \leq \frac{k_3 n W_A}{H^3 \log^2{n}}$, where $k_3$ is a constant.
\done

We then derive the upper bounds on the per-node throughput capacity under Interference Condition.

\begin{proposition}
\label{prop:regime-ii}
When Interference requirement dominates, the per-node throughput capacity contributed by ad hoc communications is $\lambda_a=O\bigg(\frac{nW_A}{C_A^{\frac{1}{2}} H^3 \log^{\frac{3}{2}} n}\bigg)$.
\end{proposition}
\textbf{Proof.} 
We present a proof of the bound in Appendix A. \done

Before proving the upper bounds on the throughput capacity under the destination-bottleneck condition, we need to bound the number of flows towards a node under the $H$-max-hop routing scheme. Specifically, we have the following result.

\begin{lemma}
\label{lemma:dh}
The maximum number of flows towards a node under the $H$-max-hop routing scheme is 
$D_H(n)=\Theta\bigg(\frac{\log(H^2 \log n)}{\log \log(H^2 \log n)}\bigg)$ \textit{w.h.p.}
\end{lemma}
\textbf{Proof.} 
Let $N_i (1\leq i \leq n)$ be a random variable defined as follows:
\begin{displaymath}
N_i=\left\{
\begin{array}{ll}
1 & \textrm{ if source node $i$ transmits to its destination node;} \nonumber\\
0 & \textrm{ otherwise.} \nonumber\\
\end{array}
\right.
\end{displaymath}
Let $N_t$ be a random variable representing the total number of source nodes transmitting in ad hoc mode. We have $N_t=\sum_{i=1}^n N_i$. Thus, the expected number of source nodes transmitting in ad hoc mode is:
\begin{displaymath}
E(N_t)=E\bigg(\sum_{i=1}^n N_i\bigg)=\sum_{i=1}^n E(N_i).
\end{displaymath}

Since $f(N_i=1)=P(AH)=\pi H^2 r^2(n)$ and $r(n)$ needs to be $\Theta(\sqrt{\log n/n})$ to ensure that the network
is connected, we have $E(N_i)=1\cdot \pi H^2 r^2(n)+0 \cdot (1-\pi H^2 r^2(n))=\pi H^2 r^2(n)$, i.e., $E(N_i)=\Theta(\pi H^2 \frac{\log n}{n})$. Therefore, $E(N_t)=n \cdot \pi H^2 \frac{\log n}{n} = \pi H^2 \log n$.

Recall the Chernoff bounds \cite{Motwani:1995},  we have
\begin{itemize}
\item for any $\delta >0$, $P(N_t > (1+\delta)\pi H^2 \log n) < \left({{e^{\delta}} \over {(1+\delta)^{(1+\delta)}}}\right)^{\pi H^2 \log n};$
\item for any $0< \delta <1$, $P(N_t < (1-\delta)\pi H^2 \log n) < e^{-\pi H^2 \log n \cdot \delta^2 /2}.$
\end{itemize}

In summary, for any $0 < \delta <1$, we can obtain $P(|N_t - \pi H^2 \log n| > \delta \pi H^2 \log n ) < e^{- \varepsilon \pi H^2 \log n}$, where $\varepsilon >0$. Thus, when $n \rightarrow \infty$, the total number of source nodes transmitting in ad hoc mode is $\Theta(H^2 \log n)$ \textit{w.h.p.}

In a random network, each source node can randomly choose its destination. The traffic for a source-destination pair is denoted as a \textit{flow}. Thus, it is very likely that a node will be the destination of multiple flows. It is proved in \cite{raab98:balls} that the maximum number of flows towards any given node in a random network with $N$ nodes, denoted by $D(N)$, is upper bounded by $\Theta\big(\frac{\log N}{\log{\log{N}}}\big)$, \textit{w.h.p.}

Combining the two results (by replacing $N=H^2\log n$) leads to the above result.
\done

We then prove the upper bounds on the per-node throughput capacity under Destination-bottleneck Condition.
\begin{proposition}
\label{prop:regime-iii}
When Destination-bottleneck requirement dominates, the per-node throughput capacity contributed by ad hoc communications is $\lambda_a=O\bigg(\frac{n^{\frac{3}{2}} \log \log (H^2 \log n) W_A}{C_A H^3 \log^{\frac{3}{2}}n \cdot \log(H^2 \log n)}\bigg)$.
\end{proposition}
\textbf{Proof.}
Since each node has one interface that can support at most $\frac{W_A}{C_A}$ and Since each node has at most $D_H(n)$ flows under the $H$-max-hop routing scheme, the data rate of the minimum rate flow is at most $\frac{W_A}{C_A D_H(n)}$, where $D_H(n)$ is bounded by $\Theta\Big(\frac{\log(H^2 \log n)}{\log \log(H^2 \log n)}\Big)$ by Lemma \ref{lemma:dh}. After calculating all the data rates at each node times with the traversing distance, we have $n \cdot P(AH) \cdot \lambda_a \cdot \overline{h} \cdot r(n) \leq \frac{W_A n}{C_A D_H(n)} \cdot 1$.

We then have
\begin{displaymath}
\lambda_a \leq \frac{W_A}{C_A D_H(n) P(AH) \overline{h} r(n)} \leq \frac{W_A}{C_A \pi H^3 r^3(n) \cdot \frac{\log (H^2 \log n)}{\log \log(H^2 \log n)}}.
\end{displaymath}
This is because $\overline{h}=\Theta(H)$ and $P(AH) = \pi H^2 r^2(n)$ are derived in Lemma \ref{lemma:H} and Eq. (\ref{eqn:prob_ah}) in the proof of Proposition \ref{prop:regime-i}, respectively. 

Since $r(n)=\Theta\Big(\sqrt{\frac{\log n}{n}}\Big)$ as proved in \cite{Gupta:Kumar}, we then have 
\begin{displaymath}
\lambda_a \leq \frac{W_A n^{\frac{3}{2}} \cdot \log \log(H^2 \log n)}{C_A H^3 \log^{\frac{3}{2}n} \cdot \log(H^2 \log n)}.
\end{displaymath}\done
 
Finally, we prove the upper bounds on the per-node throughput capacity under Interface-bottleneck Condition.
\begin{proposition}
\label{prop:regime-iv}
When Interface-bottleneck requirement dominates, the per-node throughput capacity contributed by ad hoc communications is $\lambda_a=O(\frac{W_A}{C_A})$.
\end{proposition}
\textbf{Proof.}
In an \textit{MC-IS} network, each node is equipped with only one interface, which can support at most $\frac{W_A}{C_A}$ data rate. Thus, $\lambda_a$ is also upper bounded by $\frac{W_A}{C_A}$. Note that this result holds for any network settings. 
\done

\subsection{Constructive Lower Bounds on Network Capacity Contributed by Ad Hoc Communications}
\label{sec:lower-adhoc}
We then derive the lower bound on the network capacity by constructing a network with the corresponding routing scheme and scheduling scheme when each requirement is considered. The derived orders of the lower bounds are the same as the orders of the upper bounds, meaning that the upper bounds are tight. In particular, we first divide the plane into a number of equal-sized cells. The size of each cell is properly chosen so that each cell has $\Theta(n a(n))$ nodes, where $a(n)$ is the area of a cell (Section \ref{sec:cell}). We then design a routing scheme to assign the number of flows at each node evenly (Section \ref{sec:routingscheme}). Finally, we design a \textit{Time Division Multiple Access (TDMA)} scheme to schedule the traffic at each node (Section \ref{sec:schedulingscheme}).

\subsubsection{Cell Construction}
\label{sec:cell}

\begin{figure}[t!]
\centering
 \includegraphics[width=4cm]{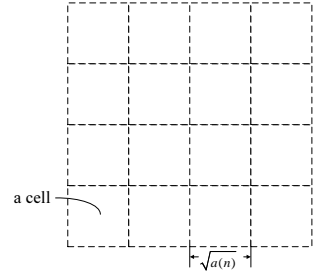}
  \caption{Plane divided into a number of cells and each with area $a(n)$.}
 \label{fig:cells}
\end{figure}

We divide the plane into $1/a(n)$ equal-sized cells and each cell is a square with area of $a(n)$, as shown in Fig. \ref{fig:cells}. The cell size of $a(n)$ must be carefully chosen to fulfill the three requirements, i.e., the connectivity requirement, the interference requirement and the destination-bottleneck requirement. In particular, similar to \cite{Kyasanur:mobicom2005}, we set $a(n)=\min\bigg\{\max\bigg\{\frac{100 \log n}{n}, \frac{ \log^{\frac{3}{2}}n}{C_A^{\frac{1}{2}} n}\bigg\}, \frac{\log^{\frac{3}{2}} n \cdot \log(H^2 \log n)}{n^{\frac{3}{2}} \cdot \log \log(H^2 \log n)}\bigg\}$. Note that the interface-bottleneck requirement is independent of the size of a cell.

The maximum number of nodes in a cell can be upper bounded by the following lemma.
\begin{lemma}
\label{lemma:no_nodes}
If $a(n)>\frac{50\log n}{n}$, then each cell has $\Theta(n(a(n))$ nodes \textit{w.h.p.}.
\end{lemma}
\textbf{Proof.}
Please refer to \cite{Kyasanur:mobicom2005}.
\done

We next check whether all the above values of $a(n)$ are properly chosen such that each cell has $\Theta(n(a(n))$ nodes \textit{w.h.p.} when $n$ is large enough (i.e., Lemma \ref{lemma:no_nodes} is satisfied). It is obvious that $\frac{100 \log n}{n} > \frac{50 \log n}{n}$ and $\frac{\log^{\frac{3}{2}}n}{C_A^{\frac{1}{2}} n} > \frac{50 \log n}{n}$ since we only consider $C_A$ in Connectivity Condition and Interference Condition. Besides, $\frac{\log^{\frac{3}{2}} n \cdot \log(H^2 \log n)}{n^{\frac{3}{2}} \cdot \log \log(H^2 \log n)}$ is also greater than $\frac{50 \log n}{n}$ with large $n$ since $\frac{\log(H^2 \log n)}{\log \log(H^2 \log n)}>1$ and $\frac{\log ^{\frac{3}{2}}n}{n^{\frac{3}{2}}}> \frac{50 \log n}{n}$ when $n$ is large enough.

Besides, the number of interfering cells around a cell is bounded by a constant, given by Lemma \ref{lemma:interfering_cells} as follows.
\begin{lemma}
\label{lemma:interfering_cells}
Under the interference model, the number of interfering cells of any given cell is bounded by a constant $k_5$, which is independent of $n$.
\end{lemma}
\textbf{Proof.} 
The detailed proof is stated in Appendix B. \done

\subsubsection{Routing Scheme}
\label{sec:routingscheme}
To assign the flows at each node evenly, we design a routing scheme consists of two steps: (1) Assigning sources and destinations and (2) Assigning the remaining flows in a balanced way.

In Step (1), each node is the originator of a flow and each node is the destination of at most $D_H(n)$ flows, where $D_H(n)$ is defined in Lemma \ref{lemma:dh}. Thus, after Step (1), there are at most $1+D_H(n)$ flows. 

We denote the straight line connecting a source S to its destination D as an S-D lines. In Step (2), we need to calculate the number of S-D lines (flows) passing through a cell so that we can assign them to each node evenly. Specifically, we have the following result.

\begin{lemma}
\label{lemma:lines}
The number of S-D lines passing through a cell is bounded by $O(n H^3 (a(n))^2)$.
\end{lemma}
\textbf{Proof.}
The detailed proof is stated in Appendix C. \done

As shown in Lemma \ref{lemma:no_nodes}, there are $\Theta(n \cdot a(n))$ nodes in each cell. Therefore, Step (2) will assign to any node at most $O\Big(\frac{n H^3 (a(n))^2}{n \cdot a(n)}\Big)= O(H^3 a(n))$ flows. Summarizing Step (1) and Step (2), there are at most $f(n)=O(1+H^3 a(n)+D_H(n))$ flows at each node. On the other hand, $H^3 a(n)$ dominates $f(n)$ since $H>1$ and $a(n)$ is asymptotically larger than $D_H(n)$ when $n$ is large enough. Thus, we have $f(n)=O(H^3 a(n))$.

\subsubsection{Scheduling Transmissions}
\label{sec:schedulingscheme}
We next design a scheduling scheme to transmit the traffic flows assigned in a \textit{routing scheme}. Any transmissions in this network must satisfy the two additional constraints simultaneously: 1) each interface only allows one transmission/reception at the same time, and 2) any two transmissions on any channel should not interfere with each other. 

\begin{figure}[t!]
\centering
 \includegraphics[width=7cm]{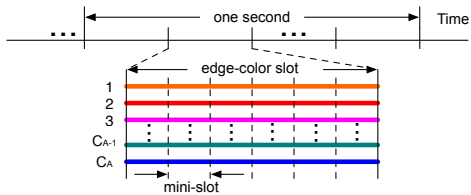}
\caption{TDMA transmission schedule}
\label{fig:schedule}
\end{figure}

We propose a TDMA scheme to schedule transmissions that satisfy the above two constraints. Fig. \ref{fig:schedule} depicts a schedule of transmissions on the network. In this scheme, one second is divided into a number of \textit{edge-color} slots and at most one transmission/reception is scheduled at every node during each edge-color slot. Hence, the first constraint is satisfied. Each edge-color slot can be further split into smaller \textit{mini-slots}. In each mini-slot, each transmission satisfies the above two constraints. 

Then, we describe the two time slots as follows. 

(i) \emph{Edge-color slot}: First, we construct a routing graph in which vertices are the nodes in the network and an edge denotes transmission/reception of a node. In this construction, one hop along a flow is associated with one edge in the routing graph. In the routing graph, each vertex is assigned with $f(n)=O(H^3 a(n))$ edges. It is shown in \cite{Kyasanur:mobicom2005,Douglas:2001} that this routing graph can be edge-colored with at most $O(H^3 a(n))$ colors. We then divide one second into $O(H^3 a(n))$ edge-color slots, each of which has a length of $\Omega(\frac{1}{H^3 a(n)})$ seconds and is stained with a unique edge-color. Since all edges connecting to a vertex use different colors, each node has at most one transmission/reception scheduled in any edge-color time slot. 
	
(ii) \emph{Mini-slot}: We further divide each edge-color slot into mini-slots. Then, we build a schedule that assigns a transmission to a node in a mini-slot within an edge-color slot over a channel. We construct an \textit{interference graph} in which each vertex is a node in the network and each edge denotes the interference between two nodes. We then show as follows that the interference graph can be vertex-colored with $k_7(n a(n))$ colors, where $k_7$ is a constant defined in \cite{Kyasanur:mobicom2005}.
\begin{lemma}
The interference graph can be vertex-colored with at most $O(n a(n))$ colors.
\end{lemma}
\textbf{Proof.}
By Lemma \ref{lemma:interfering_cells}, every cell has at most a constant number of interfering cells. Besides, each cell has $\Theta(n a(n))$ nodes by Lemma \ref{lemma:no_nodes}. Thus, each node has at most $O( n a(n))$ edges in the interference graph. It is shown that a graph of degree at most $k_0$ can be vertex-colored with at most $k_0+1$ colors \cite{Kyasanur:mobicom2005} \cite{Douglas:2001}. Hence, the interference graph can be vertex-colored with at most $O(n a(n))$ colors. \done

We need to schedule the interfering nodes either on different channels, or at different mini-slots on the same channel since two nodes assigned the same vertex-color do not interfere with each other, while two nodes stained with different colors may interfere with each other. We divide each edge-color slot into $\left\lceil \frac{k_7 n a(n)}{C_A}\right\rceil$ mini-slots on every channel, and assign the mini-slots on each channel from 1 to $\left\lceil \frac{k_7 n a(n)}{C_A}\right\rceil$. A node assigned with a color $s$, $1\leq s \leq  k_7 n a(n)$, is allowed to transmit in mini-slot $\left\lceil \frac{s}{C_A} \right\rceil$ on channel $(s \textrm{ mod } C_A) +1$.

We next prove the constructive lower bounds of the capacity. 
\begin{proposition}
\label{prop:lower-adhoc}
The achievable per-node throughput capacity $\lambda_a$ contributed by ad hoc communications is as follows.
\begin{enumerate}
\item[1)] When Connectivity requirement dominates, $\lambda_a$ is $\Omega\Big(\frac{nW_A}{H^3 \log^2 n }\Big)$ bits/sec;
\item[2)] When Interference requirement dominates, $\lambda_a$ is $\Omega\Big(\frac{nW_A}{H^3 C_A^{\frac{1}{2}}\log^{\frac{3}{2}}n}\Big)$ bits/sec;
\item[3)] When Destination-bottleneck requirement dominates, $\lambda_a$ is $\Omega\bigg(\frac{n^{\frac{3}{2}} \log \log (H^2 \log n) W_A}{C_A H^3 \log^{\frac{3}{2}}n \cdot \log(H^2 \log n)}\bigg)$ bits/sec;
\item[4)] When Interface-bottleneck requirement dominates, $\lambda_a$ is $\Omega\big(\frac{W_A}{C_A}\big)$.
\end{enumerate}
\end{proposition}
\textbf{Proof.}
Since each edge-color slot with a length of $\Omega\big(\frac{1}{H^3 a(n)}\big)$ seconds is divided into $\left\lceil \frac{k_7 n a(n)}{C_A}\right\rceil$ mini-slots over every channel, each mini-slot has a length of $\Omega\Big(\big(\frac{1}{H^3 a(n)}\big)/\left\lceil \frac{k_7 n a(n)}{C_A}\right\rceil\Big)$ seconds. Besides, each channel can transmit at the rate of $\frac{W_A}{C_A}$ bits/sec, in each mini-slot, $\lambda_a=\Omega\Big(\frac{W_A}{C_A H^3 a(n) \cdot \left\lceil \frac{k_7 n a(n)}{C_A}\right\rceil}\Big)$ bits can be transported. Since $\left\lceil \frac{k_7 n a(n)}{C_A}\right\rceil \leq \frac{k_7 n a(n)}{C_A} +1$, we have $\lambda_a=\Omega\Big(\frac{W_A}{k_7 H^3 a^2(n) n + H^3 a(n) C_A}\Big)$ bits/sec. Thus, $\lambda_a=\Omega\Big(MIN_O\Big(\frac{W_A}{H^3 a^2(n) n},\frac{W_A}{H^3 a(n) C_A}\Big)\Big)$ bits/sec \footnote{$MIN_O(f(n),g(n))$ is equal to $f(n)$ if $f(n)=O(g(n))$; otherwise, it is equal to $g(n)$.}. 

Recall that $a(n)$ is set to $\min\Big\{\max\Big\{\frac{100 \log n}{n}, \frac{ \log^{\frac{3}{2}}n}{C_A^{\frac{1}{2}} n}\Big\}, \frac{\log^{\frac{3}{2}} n \cdot \log(H^2 \log n)}{n^{\frac{3}{2}} \cdot \log \log(H^2 \log n)}\Big\}$. Substituting the three values to $\lambda_a$, we have the results 1), 2) and 3). Besides, each interface can transmit or receive at the rate of $\frac{W_A}{C_A}$ bits/sec. Thus, $\lambda_a=\Omega\big(\frac{W_A}{C_A}\big)$, which is the result 4).
\done

\subsection{Aggregate Throughput Capacity}
\label{sec:summary:ah}

The upper bounds proved in Propositions \ref{prop:regime-i}, \ref{prop:regime-ii}, \ref{prop:regime-iii} match with the lower bounds proved in Proposition \ref{prop:lower-adhoc}, implying that our bounds are quite tight. Besides, it is shown in \cite{panli:jsac09} that the total traffic of ad hoc communications is $n \pi H^2 r^2(n) \lambda_a$. Combining Propositions \ref{prop:regime-i}, \ref{prop:regime-ii}, \ref{prop:regime-iii}, and \ref{prop:lower-adhoc} with the total traffic leads to the following theorem.

\begin{theorem}
\label{theorem:ad_hoc}
The aggregate throughput capacity of the network contributed by ad hoc communications is
\begin{enumerate}
\item[1)] When Connectivity requirement dominates, $T_A$ is $\Theta(\frac{n W_A}{H \log n })$ bits/sec.
\item[2)] When Interference requirement dominates, $T_A$ is $\Theta(\frac{n W_A}{C_A^{\frac{1}{2}} H \log^{\frac{1}{2}}n})$ bits/sec.
\item[3)] When Destination-bottleneck requirement dominates, $T_A$ is $\Theta(\frac{n^{\frac{3}{2}} \log \log(H^2 \log n)W_A}{C_A H \log^{\frac{1}{2}}n \cdot \log(H^2 \log n)})$ bits/sec.
\item[4)] When Interface-bottleneck requirement dominates, $T_A$ is $\Theta(H^2 \log n \cdot \frac{W_A}{C_A})$ bits/sec.
\end{enumerate}
\end{theorem}

\section{Capacity Contributed by Infrastructure Communications}
\label{sec:infra}

In this section, we analyze the network capacity contributed by infrastructure communications. Specifically, we derive the upper bounds of the capacity in Section \ref{sec:upper-infra} and give the constructive lower bounds of the capacity in Section \ref{sec:lower-infra}. We give the aggregate capacity contributed by infrastructure communications in Section \ref{sec:infra:summary}. Note that our proposed \emph{MC-IS} networks have the multiple interfaces at a base station, compared with a single interface at a base station in \emph{SC-IS} networks. As a result, our \emph{MC-IS} networks have a better performance than \emph{SC-IS} networks though the derivations are also more complicated than those of \emph{SC-IS} networks. Finally, Section \ref{sec:proof} gives the proof of Theorem \ref{th:throughput} and Theorem \ref{th:delay}.

\subsection{Upper Bounds of Network Capacity Contributed by Infrastructure Communications}
\label{sec:upper-infra}

We derive the upper bounds of the throughput capacity contributed by infrastructure communications as follows.

\begin{proposition}
\label{prop:upper_infra}
Under the $H$-max-hop routing scheme, the throughput capacity contributed by infrastructure communications, 
denoted by $T_I$, is:
\begin{enumerate}
\item[(1)] When $C_I \leq m$, $T_I=O(b W_I)$. 
 
\item[(2)] When $C_I > m$, $T_I=O(b \frac{m}{C_I} W_I)$.
\end{enumerate}
\end{proposition}
\textbf{Proof.}
Since each packet transmitted in the infrastructure mode will use both the uplink and the downlink communications, we only count once for the throughput capacity.

\textit{Case (1) when $C_I \leq m$.}
It is obvious that the $m$ interfaces at each base station can support at most $W_I$ bandwidth. In other words, the $C_I$ channels are fully utilized by the $m$ interfaces. Counting all the $b$ base stations, we have $T_I=O(bW_I)$.

\textit{Case (2) when $C_I > m$.}
When the number of interfaces is smaller than the number of channels, not all the $C_I$ channels are fully used. In fact, at most $m$ channels can be used at a time. Besides, each channel can support at most $\frac{W_I}{C_I}$ bits/sec. Thus, each base station can support at most $\frac{m}{C_I}W_I$ bits/sec. Counting all the $b$ base stations, we have $T_I=O(b\frac{m}{C_I}W_I)$.
\done

\subsection{Constructive Lower Bounds of Network Capacity Contributed by Infrastructure Transmissions}
\label{sec:lower-infra}
The lower bounds are proved by constructing a routing scheme and a transmission scheduling scheme on a regular-tessellated BS network. The derived orders of the lower bounds are the same as the orders of the upper bounds, implying that the upper bounds are tight.

\subsubsection{BS-Cell Construction by Regular Tessellation}
There are $b$ base stations regularly placed in the plane dividing the plane into a number of equal-sized \textit{BS-cells}. Note that the size of each \textit{BS-cell} may not be necessarily equal to the size of a \textit{cell}. Besides, Lemma \ref{lemma:interfering_cells} still holds even if the base stations are regularly placed in the plane. So, the number of interfering \textit{BS-cells} is also bounded by a constant, denoted by $k_8$, which is also independent of $b$.

\subsubsection{Routing and Scheduling Schemes}
\begin{figure}[t]
\centering
\includegraphics[width=3.4cm]{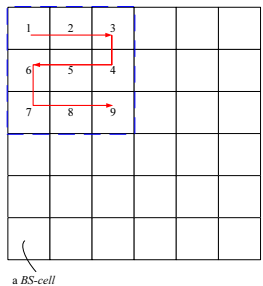}
\caption{{An example of the TDMA transmission schedule, in which each \textit{BS-cell} in a cluster becomes active every $9$ time slots.}}
\label{fig:tdma-cells}
\end{figure}

The routing scheme for the infrastructure traffic is simple, i.e., to forward the traffic to
a base station (uplink) and to forward the traffic from a base station (downlink). We propose the following TDMA scheduling scheme $\Sigma_1$ to schedule the \textit{BS-cells} to be active in a round-robin fashion. 
\begin{itemize}
\item[(1)] Divide the plane into $b$ equal-sized \textit{BS-cells}.
\item[(2)] We group the $b$ \textit{BS-cells} into a number of clusters. Each cluster has $(k_8+1)$ \textit{BS-cells}. We then split the transmission time into a number of time frames. Each frame consists of $(k_8+1)$ time slots that correspond to the number of \textit{BS-cells} in each cluster. In each time slot, one \textit{BS-cell} within each cluster becomes active to transmit and the \textit{BS-cells} in each cluster take turns to be active. For example, all the clusters follow the same 9-TDMA transmission scheduling scheme, as shown in Fig. \ref{fig:tdma-cells}.
\end{itemize}

\begin{proposition}
\label{prop:lower_infra}
Under the TDMA scheme $\Sigma_1$, the throughput capacity $T_I$, is:
\begin{enumerate}
\item[(1)] When $C_I \leq m$, $T_I=\Omega(b W_I)$. 
 
\item[(2)] When $C_I > m$, $T_I=\Omega(b \frac{m}{C_I} W_I)$.
\end{enumerate}
\end{proposition}
\textbf{Proof.}
Since each packet transmitted in the infrastructure mode will use both the uplink and the downlink,
we only count once for throughput capacity.

\textit{Case (1) when $C_I \leq m$}:
Under TDMA scheme $\Sigma_1$, each \textit{BS-cell} is active to transmit every $(k_8+1)$ time slots. When a \textit{BS-cell} is active, there are at most $C_I$ channels available to use. Thus, the total bandwidth of $W_I$ of those $C_I$ channels are fully used. Thus, the per-cell throughput $\lambda_i$ is lower bounded by $\frac{W_I}{k_8+1}$. Counting all the $b$ base stations, we have $T_I=\Omega(\frac{b W_I}{k_8+1})$.

\textit{Case (2) when $C_I > m$}:
Similarly, each \textit{BS-cell} is active to transmit every $(k_8+1)$ time slots in case (2). But, when a \textit{BS-cell} is active, only $m$ channels available at a time and each channel can support at most $\frac{W_I}{C_I}$ data rate. Thus, the per-cell throughput $\lambda_i$ is lower bounded $\frac{m W_I}{C_I (k_8+1)}$. Counting all the $b$ base stations, we have $T_I=\Omega(\frac{b m W_I}{C_I(k_8+1)})$.
\done

\subsection{Aggregate Throughput Capacity}
\label{sec:infra:summary}
After combining Proposition \ref{prop:upper_infra} and Proposition \ref{prop:lower_infra}, 
we have the following theorem.

\begin{theorem}
\label{theorem:infra}
The aggregate throughput capacity of the network contributed by infrastructure communications is
\begin{enumerate}
\item[(1)] When $C_I \leq m$, $T_I=\Theta(b W_I)$. 
\item[(2)] When $C_I > m$, $T_I=\Theta(b \frac{m}{C_I} W_I)$.
\end{enumerate}
\end{theorem}

It is shown in Theorem \ref{theorem:infra} that the optimal throughput capacity contributed by infrastructure communications $T_I=\Theta(b W_I)$ is achieved when $C_I \leq m$. Generally, we have $C_I = m$. If $C_I \neq m$, some interfaces are idle and wasted. It implies that to maximize $T_I$, we shall assign a dedicated interface per channel at each base station so that all the $C_I$ channels can be fully utilized. However, it is not true that we always have $C_I = m$ since the radio spectrum becomes scarce \cite{Akyildiz:2006} and there may be fewer channels than the interfaces. We will give a discussion on this issue in Section \ref{sec:discussion}.

\subsection{Proof of Theorem \ref{th:throughput} and Theorem \ref{th:delay}}
\label{sec:proof}

We finally give the proof of Theorem \ref{th:throughput} and Theorem \ref{th:delay} as follows:

\textbf{Proof of Theorem \ref{th:throughput}}

We first have the aggregate throughput capacity $T = T_A + T_I$, where $T_A$ is the aggregate capacity contributed by ad hoc communications and $T_I$ is the aggregate capacity contributed by infrastructure communications given by given by Theorem \ref{theorem:ad_hoc} and Theorem \ref{theorem:infra}, respectively. Since there are at most $n$ nodes in the network, we then divide $T$ by $n$ and finally have the results in Theorem \ref{th:throughput}. This completes the proof.
\done

We then derive the average delay of an \textit{MC-IS} network contributed by ad hoc communications and infrastructure communications as follows.

\textbf{Proof of Theorem \ref{th:delay}}

We first derive the bound on the delay when the packets are transmitted in the infrastructure mode. As shown in \cite{panli:jsac09}, the average delay for the packets transmitted in the infrastructure mode in an \textit{SC-IS} network is bounded by $\Theta(c)$, where $c$ is a constant depending on the transmitting capability of the base station. Different from an \textit{SC-IS} network, where each base station is equipped with a single interface supporting at most one transmission at a time, each base station in an \textit{MC-IS} network can support $\min\{C_I,m\}$ simultaneous transmissions at a time. This is because when $C_I \leq m$, a base station with $m$ interfaces can support at most $C_I$ simultaneous transmissions; when $C_I > m$, a base station with $m$ interfaces can support at most $m$ simultaneous transmissions. Thus, the average delay for the packets transmitted in the infrastructure mode in an \textit{MC-IS} network is bounded by $\Theta(\frac{c}{\min\{C_I,m\}})$.

We then derive the bound on the delay when the packets are transmitted in ad hoc mode. If the packets are transmitted in the ad hoc mode, the expectation of the number of hops $\overline{h}$ under the $H$-max-hop routing strategy is bounded by $\Theta(H)$ as proved by Lemma \ref{lemma:H}. Since the time spent by a packet at each relay is bounded by a constant number $c_1$, the average delay is of the same order as the average number of hops, i.e., $D=c_1 \cdot \overline{h} = \Theta(H)$. 

It is shown in the proof of Lemma \ref{lemma:dh} that the number of transmitters in the ad hoc mode is $\pi H^2 \log n$ \textit{w.h.p.} Then the number of transmitters in the infrastructure mode is $(n - \pi H^2 \log n)$ \textit{w.h.p.} After applying the aforementioned analysis, we have the average delay of all packets $D = \Theta\Big(\frac{\pi H^2 \log n \cdot H + (n - \pi H^2 \log n) \cdot \frac{c}{\min\{C_I,m\}}}{n}\Big)$. Note that $\frac{n - \pi H^2 \log n}{n}$ is bounded by $\Theta(1)$. Thus, $D= \Theta\big(\frac{H^3 \log n}{n}\big) + \Theta\big(\frac{c}{\min\{C_I,m\}}\big)$. 
\done

\section{Discussions and Implications}
\label{sec:discussion}
In this section, we first extend our analysis to the scenarios of using directional antennas in \emph{MC-IS} networks in Section \ref{subsec:dir}. Note that our analysis is non-trivial since the existing analytical models such as \emph{MC-AH} networks, \emph{SC-IS} networks and even our \emph{MC-IS} networks cannot be directly used in the extended \emph{MC-IS} networks because the interference model is significantly different from those existing ones. We then discuss the impacts of mobility models in Section \ref{subsec:mobility}. Finally, we present the implications of our \emph{MC-IS} networks in Section \ref{subsec:implications}.

\subsection{Using Directional Antennas in MC-IS networks}
\label{subsec:dir}

Conventional wireless networks assume that each node is equipped with an \emph{omni-directional} antenna, which radiates radio signals in all directions including some undesired directions. Recent studies such as \cite{Su:2003,panli:capacity11} show that applying directional antennas instead of omni-directional antennas to wireless networks can greatly improve the network capacity. The performance improvement mainly owes to the reduction in the interference from undesired directions since directional antennas concentrate radio signals on the desired directions. Although directional antennas have numerous advantages, the bulky size and the impacts of directionality also restrict the application of directional antennas to conventional wireless networks. However, with the evolution of wireless communication technologies, these challenging issues will finally be solved. In fact, a directional antenna has become a necessity in order to compensate for the tremendous signal attenuation in millimeter-wave (mmWave) communication systems, which is a very promising solution for the next generation communication systems (5G) \cite{Rappaport:Access2013}. It is feasible to deploy directional antennas at both base stations and mobile devices in mmWave communication systems since their size will be quite compact due to the fact that the antenna size is inversely proportional to the radio frequency (the frequency band is ranging from 30GHz to 300GHz in mmWave communication systems \cite{JQiao:IEEECommMag15}).

\begin{figure}[t]
\centering
\includegraphics[width=6cm]{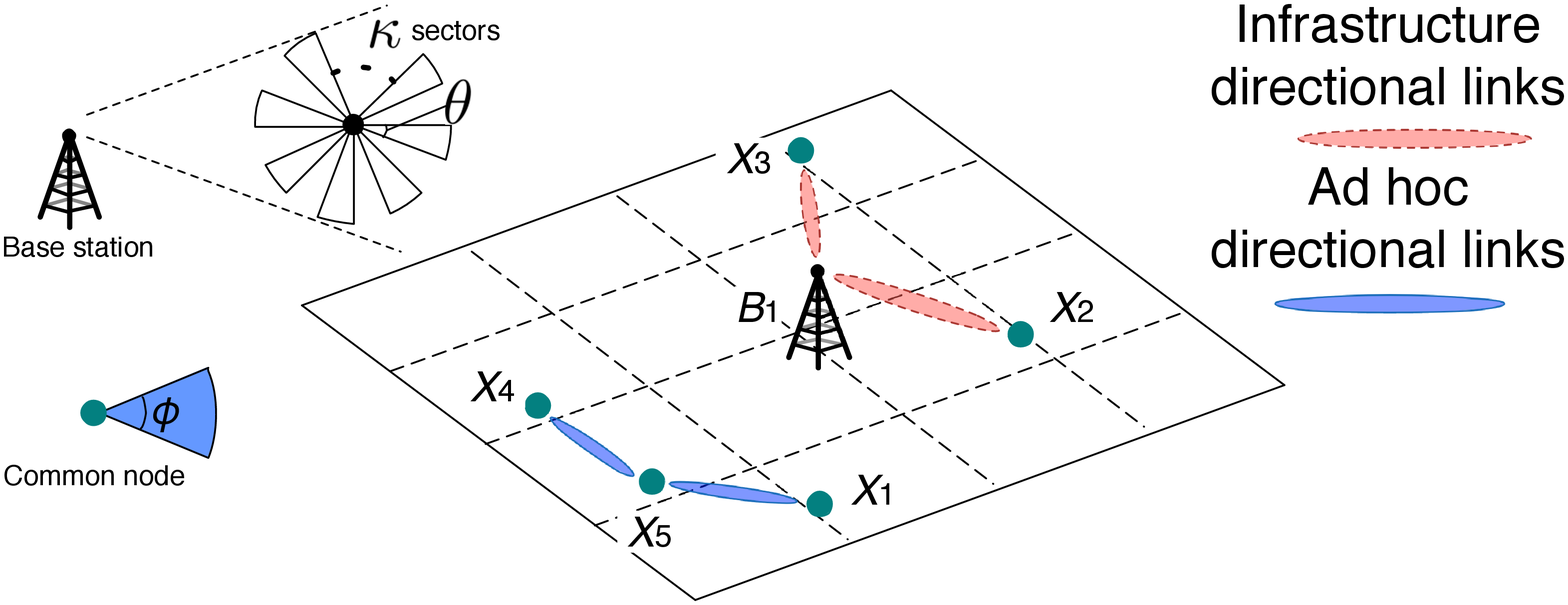}
\caption{Network topology of an \emph{MC-IS-DA} network in a \emph{BS-cell}}
\label{fig:MC-IS-DA}
\end{figure}

We extend our analysis on an \emph{MC-IS} network with omni-directional antennas (in the previous part of this paper) to that with directional antennas. In particular, we name an \emph{MC-IS} network equipped with directional antennas as an \emph{MC-IS-DA} network. Fig. \ref{fig:MC-IS-DA} shows an example of \emph{MC-IS-DA} networks, in which each base station is equipped with multiple directional antennas and each common node is equipped with a single directional antenna. Similar to an \emph{MC-IS} network, there are two types of communications in an \emph{MC-IS-DA} network: {\it ad hoc communications} between common nodes and {\it infrastructure communications} between a common node and a base station. Differently, both ad hoc communications and infrastructure communications in an \emph{MC-IS-DA} network consist of \emph{directional communication links} only.

In this paper, we consider a \emph{flat-top} antenna model that is typically used in previous works \cite{Su:2003,hndai:infocom2008,Nitsche:IEEEComMag14}. In particular, sidelobes and backlobes are ignored in this directional antenna model. This is because the sidelobes/backlobes are so small that the impacts of them can be ignored when the main beamwidth is small (e.g., 30$\,^{\circ}$ in \cite{Ramanathan:2001}). Besides, smart antennas often have null capability that can almost eliminate the sidelobes and backlobes \cite{Balanis:2005}. Our antenna model assumes that a directional antenna gain is within a specific angle, i.e., the beamwidth of the antenna, which is ranging from 0 to $\pi$. The gain outside the beamwidth is assumed to be zero. In our \emph{MC-IS} network, each common node is mounted with a single interface, which is equipped with a directional antenna with beamwidth $\phi$. Each base station is mounted with $m$ interfaces, each of which is equipped with a directional antenna with beamwidth $\theta$, where each directional antenna at each base station is identical. Note that the beamwidth $\phi$ of an antenna at a common node is not necessarily equal to the beamwidth $\theta$ of that at a base station. 

\emph{1. Capacity of an MC-IS-DA network}

We first derive the capacity of an \emph{MC-IS-DA} network. In particular, the capacity of an \emph{MC-IS-DA} network contributed by infrastructure communications is the same as that of a typical \emph{MC-IS} network. With regard to the capacity contributed by ad hoc communications, we need to extend our analysis in Section \ref{sec:ad-hoc} to an \emph{MC-IS-DA} network. Specifically, an \emph{MC-IS-DA} network has different capacity regions on the per-node throughput capacity $\lambda_a$, compared with an \emph{MC-IS} network. We show the main results as follows.

\begin{corollary}
\label{cor:throughput}
The per-node throughput $\lambda$ for an \textit{MC-IS-DA} network has four regions as follows.
\begin{enumerate}
\item [i)] When \textit{Connectivity Condition} is satisfied, $\lambda = \Theta\big(\frac{4\pi^2}{\phi^2} \cdot \frac{W_A}{ H \log n}\big)+\Theta\big(\min\{\frac{b}{n},\frac{bm}{nC_I}\}W_I\big)$, where $\lambda_a = \Theta\big(\frac{4\pi^2}{\phi^2} \cdot\frac{W_A}{ H \log n}\big)$ and $\lambda_i = \Theta\big(\min\{\frac{b}{n},\frac{bm}{nC_I}\}W_I\big)$;
\item [ii)] When \textit{Interference Condition} is satisfied, $\lambda = \Theta\Big(\frac{2\pi}{\phi} \cdot\frac{ W_A}{C_A^{\frac{1}{2}} H \log^{\frac{1}{2}}n}\Big)+\Theta(\min\{\frac{b}{n},\frac{bm}{nC_I}\}W_I)$, where $\lambda_a =\Theta\Big(\frac{2\pi}{\phi} \cdot \frac{ W_A}{C_A^{\frac{1}{2}} H \log^{\frac{1}{2}}n}\Big)$ and $\lambda_i = \Theta\big(\min\{\frac{b}{n},\frac{bm}{nC_I}\}W_I\big)$;
\item [iii)] When \textit{Destination-bottleneck Condition} is satisfied, $\lambda = \Theta\bigg(\frac{n^{\frac{1}{2}} \log \log(H^2 \log n)W_A}{C_A H \log^{\frac{1}{2}}n \cdot \log(H^2 \log n)}\bigg)+\Theta(\min\{\frac{b}{n},\frac{bm}{nC_I}\}W_I)$, where $\lambda_a = \Theta\bigg(\frac{n^{\frac{1}{2}} \log \log(H^2 \log n)W_A}{C_A H \log^{\frac{1}{2}}n \cdot \log(H^2 \log n)}\bigg)$ and $\lambda_i = \Theta\big(\min\{\frac{b}{n},\frac{bm}{nC_I}\}W_I\big)$;
\item [iv)] When \textit{Interface-bottleneck Condition} is satisfied, $\lambda = \Theta\bigg(H^2 \frac{\log n}{n} \cdot \frac{W_A}{C_A}\bigg)+\Theta(\min\{\frac{b}{n},\frac{bm}{nC_I}\}W_I)$, where $\lambda_a = \Theta\bigg(H^2 \frac{\log n}{n} \cdot \frac{W_A}{C_A}\bigg)$ and $\lambda_i = \Theta\big(\min\{\frac{b}{n},\frac{bm}{nC_I}\}W_I\big)$.
\end{enumerate}
\end{corollary}
\textbf{Proof.}
The detailed proof is presented in Appendix D.
\done

As shown in Corollary \ref{cor:throughput}, an \textit{MC-IS-DA} network has four capacity regions similar to an \textit{MC-IS} network. However, compared with an \textit{MC-IS} network, an \textit{MC-IS-DA} network has the higher throughput capacity than an \textit{MC-IS} network when {\it Connectivity} requirement and {\it Interference} requirement dominate. In particular, when \textit{Connectivity Condition} is satisfied, an \textit{MC-IS-DA} network has a capacity gain $\frac{4\pi^2}{\phi^2}$ over an \textit{MC-IS} network. When \textit{Interference Condition} is satisfied, an \textit{MC-IS-DA} network has a capacity gain $\frac{2\pi}{\phi}$ over an \textit{MC-IS} network. This result implies that using directional antennas in an \textit{MC-IS} network can significantly improve the capacity contributed by ad hoc communications. The capacity improvement may owe to the \emph{improved network connectivity} and the \emph{reduced interference}. One thing to note that the capacity of \textit{MC-IS-DA} network contributed by infrastructure communications $\lambda_i$ is the same as that of an \textit{MC-IS} network, implying that using directional antennas at base stations will not improve the capacity. However, our following analysis will prove that using directional antennas at base stations can significantly reduce the delay contributed by infrastructure communications.

\emph{2. Delay of an MC-IS-DA network}

Recall in Section \ref{sec:infra:summary} that we need to have $C_I \leq m$ so that the maximum throughput capacity contributed by infrastructure communications can be achieved. We usually have $C_I = m$ so that there is no waste of interfaces. It implies that we shall assign a dedicated interface per channel at each base station so that all the $C_I$ channels can be fully utilized. However, as the radio spectrum is becoming more congested and scarce \cite{Akyildiz:2006}, it is extravagant and impractical to let $C_I = m$. Thus, we extend our analysis to the case with $C_I < m$. 

We first equally divide $m$ antennas into $\kappa$ groups, each of which has $\frac{m}{\kappa}$ antennas (we also assume that $m$ is divisible by $\kappa$ though this analysis can be easily extended to the case that $m$ is not divisible by $\kappa$). Within each group, the $\frac{m}{\kappa}$ antennas are pointed to the same direction so that their beams cover each other, as shown in Fig. \ref{fig:MC-IS-DA}. We name each group of antennas as a \emph{sector}. It is obvious that each sector will cover $\theta$. There is no overlapping between any two adjacent sectors. Therefore, {\it there is no conflict between any transmissions from two adjacent sectors}. The conflict only happens between the antennas within the same sector. In order to avoid conflicts, we can assign $C_I$ channels to the conflicting transmissions within the same sector. In an \emph{MC-IS-DA} network, each base station with multiple directional antennas can support more simultaneous transmissions than that of a typical \emph{MC-IS} network. Intuitively, an \emph{MC-IS-DA} network can have a better performance than a typical \emph{MC-IS} network. In particular, we have the following result.

\begin{corollary}
\label{cor:delay}
The average delay of all packets in an extended \textit{MC-IS} network is $D=\Theta\Big(\frac{H^3 \log n}{n}\Big) + \Theta\Big(\frac{c}{\lfloor \frac{2\pi}{\theta}\rfloor \cdot C_I}\Big)$, where $D_a=\Theta\Big(\frac{H^3 \log n}{n}\Big)$ and $D_i=\Theta\Big(\frac{c}{\lfloor \frac{2\pi}{\theta}\rfloor \cdot C_I}\Big)$.
\end{corollary}
\textbf{Proof.} Since the delay contributed by ad hoc communications $D_a$ is the same as the aforementioned analysis in Section \ref{sec:proof}, we omit the analysis of $D_a$ here.

We next derive the bound on the delay when the packets are transmitted in the infrastructure mode. Since the average delay for the packets transmitted in the infrastructure mode in an \textit{SC-IS} network is bounded by $\Theta(c)$ \cite{panli:jsac09}, where $c$ is a constant depending on the transmitting capability of the base station. In an \emph{MC-IS-DA} network, each base station now has $m$ directional antennas, which can support at most $m$ simultaneous transmissions at a time. Thus, theoretically, our \emph{MC-IS-DA} network shall have $m$ times transmitting capability than that of an \textit{SC-IS} network. The remaining question is what the maximum value of $m$ is. We then derive the upper bound on $m$ as follows. 

It is obvious that $m$ is determined by $\kappa$ and the number of available channels $C_I$. Specifically, we have $m =\kappa \cdot C_I$ from our \emph{MC-IS-DA} network model. Besides, $\kappa$ is apparently upper bounded by $\lfloor \frac{2\pi}{\theta}\rfloor$ since there is no overlapping between any two sectors. Then, with each sector, there are at most $C_I$ channels that can be used. Therefore, $m = \lfloor \frac{2\pi}{\theta}\rfloor \cdot C_I$.

In summary, the average delay for the packets transmitted in the infrastructure mode in an \emph{MC-IS-ext2} network is bounded by $\Theta\Big(\frac{c}{\lfloor \frac{2\pi}{\theta}\rfloor \cdot C_I}\Big)$. Following the similar proof steps of Theorem \ref{th:delay}, we have the above result.
\done

It is shown in Corollary \ref{cor:delay} that using directional antennas at base stations in an \emph{MC-IS} network can further reduce the average delay contributed by infrastructure communications $D_i$ in the case $C_I < m$ since obviously $\lfloor \frac{2\pi}{\theta}\rfloor C_I > C_I$.  Besides, Corollary \ref{cor:delay} also shows that the narrower antenna beamwidth $\theta$ is, the lower average delay $D_i$ is. This result also implies that using directional antennas in an \emph{MC-IS} network can significantly improve the spectrum reuse. For example, suppose that we only have only one channel available, i.e. $C_I=1$, which can only be used by one omni-directional antenna in an \emph{MC-IS} network even if there are more than one antennas. However, in an \emph{MC-IS-DA} network where each base station is equipped with 12 directional antennas each with beamwidth $\frac{\pi}{6}$ (i.e., 30$\,^{\circ}$), this single channel can be simultaneously used by 12 antennas. Actually, in mmWave cellular networks, the beamwidth of a directional antenna is usually less than 15$\,^{\circ}$ \cite{Rappaport:Access2013} potentially improving the spectrum reuse better.

\subsection{Impacts of Mobility}
\label{subsec:mobility}

Multi-hop and short-ranged ad hoc communications inevitably result in the low throughput and the high delay due to the interference among multiple concurrent transmissions and the time spent on multi-hop relays. As shown in \cite{DTSE:TON2002}, to allow a mobile node to serve as the relay between the source and the destination can greatly reduce the interference and consequently lead to the higher throughput than the network without mobile relays. In \emph{MC-IS} networks, we can also employ mobile nodes to serve as the relays similar to \cite{DTSE:TON2002}. Note that the mobility can only be applied to \emph{common} nodes instead of infrastructure nodes (\emph{base stations}) since all the base stations are connected through a wired network with high bandwidth and they are usually fixed. When there is the similar assumption on the mobile model (i.e. random walk) to \cite{DTSE:TON2002}, we shall be able to derive the higher throughput capacity contributed by ad hoc communications, which shall be bounded by $\Theta(W_A)$ as suggested in \cite{DTSE:TON2002}. Since the proving techniques are similar to those in \cite{DTSE:TON2002}, we ignore the detailed derivations in this paper. 

In addition to random walk model, more realistic mobility models, such as random way-point model \cite{Sharma:ICC04} and Brownian motion model \cite{Xlin:TIT2006}. can also be used in our \emph{MC-IS} networks. It is not the focus of our paper to consider mobility in our \emph{MC-IS} networks due to the following reasons:
(1) most of existing mobility models can be directly used in ad hoc communications in our \emph{MC-IS} networks, which basically have the similar features to conventional wireless ad hoc networks; (2) introducing mobile relay nodes to the network also brings the higher delay no matter which mobility model is used, as indicated in \cite{Sharma:ICC04,Xlin:TIT2006,gamal:TIT2006}. This is because it always takes a long time for relay nodes to move from the source to the destination. How to achieve the high throughput while maintaining the low delay in \emph{MC-IS} networks is still an open problem. 

\subsection{Implications of our results}
\label{subsec:implications}

The penetration of wireless communication with mobile intelligent technologies is significantly changing our daily lives. It arises a diversity of new applications of scalable smart communication systems, e.g., wireless sensor networks (WSNs), smart grid, smart home and e-health systems \cite{YZhang:IEEENet12,YYan:TCST13}. The smart communication systems require smart devices (smart-phones, smart appliances, sensors, robots, surveillance devices) connected together. Due to the heterogeneity of devices and applications, heterogeneous traffics are generated. Take the smart grid as an example. It may require the narrower bandwidth to transmit power consumption information from smart meters to the operation center than that to transmit surveillance videos. The heterogeneity of the network performance requirements of various applications leads to the new research challenges in this area \cite{Khorov:2015}, e.g., how to improve the throughput capacity by offloading the traffic at base stations. Our \textit{MC-IS} networks \emph{provide a solution} to the above raised challenges. When there are a large number of low-volume traffics, e.g., transmitting monitored temperature information from sensors to sinks in a WSN, we need to let ad hoc communications dominate, i.e. $\lambda_a$ dominates $\lambda_i$, as implied from our aforementioned results (see Section \ref{sec:main}). On the other hand, when there are high-volume traffics, such as transmitting images or surveillance videos obtained from autonomous cameras to the controlling center of a smart grid, we need to let infrastructure communications dominate, i.e. $\lambda_i$ dominates $\lambda_a$. When there are some hybrid traffics of high-volume data and low-volume data, we need to assign ad hoc communications and infrastructure communications proportionally. There is an interesting question: how to assign the traffics to either infrastructure communications or ad hoc communications according to different bandwidth requirements of various applications. This may be left as one of future working directions in an \textit{MC-IS} network.

\emph{Device-to-Device} (D2D) communications have recently attracted great attentions since this technology can  offload the network traffic, improve the spectrum reuse and increase the throughput capacity \cite{Asadi:CST14,Tehrani:IEEECommMag14}. However, there are a number of challenges in D2D networks, such as the interference management, relay management and the resource (spectrum) allocation. D2D networks have the common features of our \emph{MC-IS} networks. For example, there are two kinds of communications in a D2D network: (i) D2D communications between devices (similar to ad hoc communications in our \emph{MC-IS} networks) and (ii) \emph{cellular} communications between devices and base stations (similar to infrastructure communications in our \emph{MC-IS} networks). Therefore, our theoretical analysis on \emph{MC-IS} networks can be used to analyze the performance of D2D networks, especially for \emph{overlaid} D2D networks\footnote{In an \emph{overlaid} D2D networks, the dedicated spectrum resources have been allocated to D2D communications and cellular communications, respectively. In an \emph{underlaid} D2D networks, both D2D communications and cellular communications are sharing the same spectrum, which nonetheless requires more complicated schemes.}. For example, in D2D networks, we can allocate $C_A$ channels for multi-hop D2D communications and allocate $C_I$ channels for cellular communications in D2D networks. The throughput and the delay of D2D networks shall have the same bounds as our \emph{MC-IS} networks. Meanwhile, our proposed $H$-max-hop routing scheme can be applied to D2D networks to solve the relay (routing) issues with multi-hop D2D communications \cite{SWJeon:icc15,YQian:TVT15} since it is more practical than conventional ad hoc routing schemes, which often traverse the whole network while our $H$-max-hop routing scheme can localize the communications within $H$ hops (which is quite practical to D2D communications).

\section{Conclusion}
\label{sec:conclusion}

In this paper, we propose a novel multi-channel wireless network with infrastructure (named an \textit{MC-IS} network), which consists of common nodes, each of which has a single interface, and infrastructure nodes, each of which has multiple interfaces. We derive the upper bounds and lower bounds on the capacity of an \textit{MC-IS} network, where the upper bounds are proved to be tight. Besides, we find that an \textit{MC-IS} network has a higher optimal capacity and the lower average delay than an \textit{MC-AH} network and an \textit{SC-AH} network. In addition, it is shown in this paper that an \textit{MC-IS} network has the same optimal capacity as an \textit{SC-IS} network while maintaining a lower average transmission delay than an \textit{SC-IS} network. Moreover, since each common node in an \textit{MC-IS} network is equipped with a single interface only, we do not need to make too many changes to the conventional ad hoc networks while obtaining high performance. We extend our analysis on an \emph{MC-IS} network equipped with \emph{omni-directional antennas} only to an \emph{MC-IS} network equipped with \emph{directional antennas} only, which are named as an \emph{MC-IS-DA} network. We show that an \emph{MC-IS-DA} network has an even lower delay of $\frac{c}{\lfloor \frac{2\pi}{\theta}\rfloor \cdot C_I}$ compared with an \emph{SC-IS} network and our \emph{MC-IS} network. For example, when $C_I=12$ and $\theta=\frac{\pi}{12}$, an \emph{MC-IS-DA} can further reduce the delay by 24 times lower that of an \emph{MC-IS} network and reduce the delay by 288 times lower than that of an \emph{SC-IS} network.

\appendices
\section{}
\textbf{Proof of Proposition \ref{prop:regime-ii}}

When Interference Condition is satisfied, the per-node throughput is limited by the interference requirement \cite{Gupta:Kumar}. Similar to \cite{Gupta:Kumar}, we assume that all nodes are synchronized. Let the average distance between a source and a destination be $\overline{l}$, which is roughly bounded by $\overline{h}\cdot r(n)$. 

In the network with $n$ nodes and under the $H$-max-hop routing scheme, there are at most $n\cdot P(\textrm{AH})$, where $P(AH)$ is the probability that a node transmits in ad hoc mode and can be calculated by Eq. (\ref{eqn:prob_ah}). Within any time period, we consider a bit $b$, $1\leq b \leq \lambda n P(AH)$. We assume that bit $b$ traverses $h(b)$ hops on the path from the source to the destination, where the $h$-th hop traverses a distance of $r(b,h)$. It is obvious that the distance traversed by a bit from the source to the destination is no less than the length of the line jointing the source and the destination. Thus, after summarizing the traversing distance of all bits, we have
\begin{displaymath}
\lambda_a \cdot n \overline{l} \cdot P(AH) \leq \sum_{b=1}^{n \lambda_a P(AH)} \sum_{h=1}^{h(b)} r(b,h).
\end{displaymath}

Let $T_h$ be the total number of hops traversed by all bits in a second and we have $T_h=\sum_{b=1}^{n \lambda_a P(AH)} h(b)$. Since each node has one interface which can transmit at most $\frac{W_A}{C_A}$, the total number of bits that can be transmitted by all nodes over all interfaces are at most $\frac{W_A n}{2C_A}$, i.e.,
\begin{equation}
\label{eqn:th0}
T_h \leq \frac{W_A n}{2C_A}.
\end{equation}

On the other hand, under the interference model, we have the following in-equation given by \cite{Gupta:Kumar}
\begin{displaymath}
\textrm{dist}(X_1-X_2)\geq \frac{\Delta}{2}(\textrm{dist}(X_3-X_4)+\textrm{dist}(X_1-X_2)),
\end{displaymath}
where $X_1$ and $X_3$ denote the transmitters and $X_2$ and $X_4$ denote the receivers. This in-equation implies that each hop consumes a disk of radiums $\frac{\Delta}{2}$ times the length of the hop.

Therefore, we have 
\begin{displaymath}
\sum_{b=1}^{n \lambda_a P(AH)} \sum_{h=1}^{h(b)} \frac{\pi \Delta^2}{4} (r(b,h))^2 \leq W_A.
\end{displaymath}

This in-equation can be rewritten as
\begin{equation}
\label{eqn:th1}
\sum_{b=1}^{n \lambda_a P(AH)} \sum_{h=1}^{h(b)} \frac{1}{T_h} (r(b,h))^2 \leq \frac{4 W_A}{\pi \Delta^2 T_h}.
\end{equation}

Since the left hand side of this in-equation is convex, we have
\begin{equation}
\label{eqn:th2}
\Bigg(\sum_{b=1}^{n \lambda_a P(AH)} \sum_{h=1}^{h(b)} \frac{1}{T_h} r(b,h)\Bigg)^2 \leq \sum_{b=1}^{n \lambda_a P(AH)} \sum_{h=1}^{h(b)} \frac{1}{T_h} (r(b,h))^2.
\end{equation}

Joining (\ref{eqn:th1})(\ref{eqn:th2}), we have
\begin{displaymath}
\sum_{b=1}^{n \lambda_a P(AH)} \sum_{h=1}^{h(b)} r(b,h) \leq \sqrt{\frac{4 W_A T_h}{\pi \Delta^2 }}.
\end{displaymath}

From (\ref{eqn:th0}), we have
\begin{displaymath}
\sum_{b=1}^{n \lambda_a P(AH)} \sum_{h=1}^{h(b)} r(b,h) \leq W_A\sqrt{\frac{2 n }{\pi \Delta^2 C_A}}.
\end{displaymath}

Besides, since $\lambda_a \cdot n \overline{l} \cdot P(AH) \leq \sum_{b=1}^{n \lambda_a P(AH)} \sum_{h=1}^{h(b)} r(b,h)$, we have
\begin{displaymath}
\lambda_a \leq \frac{W_A\sqrt{\frac{2 n}{\pi \Delta^2 C_A}}}{n \overline{l} \cdot P(AH)}
=\frac{W_A\sqrt{\frac{2 n}{\pi \Delta^2 C_A}}}{n \overline{h} r(n) \pi H^2 (r(n))^2} 
\leq \frac{W_A\sqrt{\frac{2}{\pi \Delta^2 n C_A}}}{\pi H^3 (r(n))^3}.
\end{displaymath}

Since $r(n) > \sqrt{\frac{\log n}{\pi n}}$, we finally have
\begin{displaymath}
\lambda_a \leq \frac{k_4 n W_A}{C_A^{\frac{1}{2}} H^3 \log^{\frac{3}{2}}n}.
\end{displaymath}
\done

\section{}

\begin{figure}[t]
\centering
 \includegraphics[width=2.8cm]{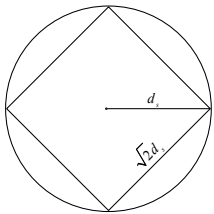}
\caption{The number of interfering cells contained in disk $D$}
\label{fig:disk}
\end{figure}

\textbf{Proof of Lemma \ref{lemma:interfering_cells}}

Consider any cell in Fig. \ref{fig:cells}. The distance between any 
transmitter and receiver within the cell can not be more than
$r_{\max}=\sqrt{2a(n)}$.

Under the interference model, a transmission can be successful if no node within distance $d_s=(1+\Delta)r_{\max}$ of the receiver transmits at the same time. Therefore, all the interfering cells must be contained within a disk $D$ as shown in Fig. \ref{fig:disk}. The number of cells contained in disk $D$ is thus bounded by:
\begin{eqnarray}
k_5 & = & \frac{(\sqrt{2}d_s)^2}{a(n)} = \frac{(\sqrt{2}(1+\Delta)r_{\max})^2}{a(n)}  = \frac{2(1+\Delta)^2 \cdot 2a(n)}{a(n)} = 4(1+\Delta)^2  \nonumber,
\end{eqnarray}
which is a constant, independent of $n$ (note that $\Delta$ is a positive constant as given in Section \ref{subsec:MC-IS}).\done

\section{}

\textbf{Proof of Lemma \ref{lemma:lines}}

\begin{figure}[t]
\centering
\includegraphics[width=3cm]{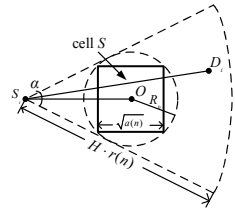}
\caption{The probability that a line $L_i$ intersects a cell $S$.}
\label{fig:lines}
\end{figure}

Consider a cell $S$, as shown in Fig. \ref{fig:lines}. It is obvious that cell $S$ is contained in a disk of radius $R_0=\frac{\sqrt{a(n)}}{2}$. Suppose $S_i$ lies at distance $x$ from the center of the disk. The angle $\alpha$ subtended at $S_i$ by the disk is no more than $\frac{k_7}{x} \cdot \sqrt{\frac{a(n)}{2}}$. It the destination node $D_i$ is not located within the sector of angle $\alpha$, the line $l_i$ cannot intersect the disk containing the cell $S$. Thus, the probability that $L_i$ intersects the disk is no more than $\frac{k_8 H^2 (r(n))^2}{x} \cdot \sqrt{\frac{a(n)}{2}}$. 

Since each source node $S_i$ is uniformly distributed in the plane of unit area, the probability density that $S_i$ is at a distance $x$ from the center of the disk is bounded by $2\pi x$. Besides, $R_0 \leq x \leq  H \cdot r(n)$. In addition, to ensure the successful transmission, the transmission range $r(n)\leq 4 R_0=\sqrt{8(a(n))}$. As a result, we have
\begin{eqnarray}
\begin{aligned}
& P\bigg(\textrm{\footnotesize $L_i$ intersects $S$ and the transmission along $L_i$ is using bandwidth $\frac{W_A}{C_A}$}\bigg) \nonumber\\
& \leq \int_{R_o}^{H\cdot r(n)} \frac{H^2}{x} \cdot ((a(n))^{\frac{3}{2}} \cdot 2\pi x \textrm{d} x \leq k_6 H^3 (a(n))^2. \nonumber 
\end{aligned}
\end{eqnarray} \done

\section{}

\textbf{Proof of Corollary \ref{cor:throughput}}

Similar to the proof of the throughput capacity of an \textit{MC-IS} network, we also derive the upper bounds and the lower bounds on the capacity of an \textit{MC-IS-DA} network. We then show that the upper bounds match with the lower bounds, implying our bounds are quite tight. After combining these results, we can prove Corollary \ref{cor:throughput}.

\emph{1. Upper bounds on Capacity of an MC-IS-DA Network}

Similarly, the capacity of an \textit{MC-IS-DA} network is also affected by all four aforementioned requirements. The main difference lies in deriving the results on Connectivity requirement and Interference requirement. Without repetitions, we only show the key proving steps in this section. 

\begin{figure}[t]
\centering
\includegraphics[width=4.5cm]{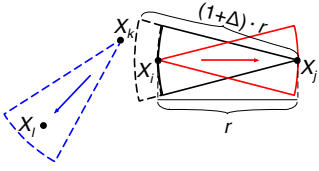}
\caption{Interference model of directional antennas.}
\label{fig:interference-dir}
\end{figure}

We first extend the \emph{interference} model in Section \ref{subsec:MC-IS} to the case of using directional antennas at both transmitters and receivers. Consider that node $X_i$ transmits to node $X_j$ over a channel. The transmission is successfully completed by node $X_j$ if no nodes within the region covered by $X_j$'s antenna beam will interfere with $X_j$'s reception. Therefore, for every other node $X_k$ simultaneously transmitting over the same channel, and the guard zone $\Delta>0$, the following condition holds
\begin{equation}
\left\{ \begin{array}{l}
\textrm{dist}(X_k,X_j)\geq(1+\Delta)\textrm{dist}(X_i,X_j)\\
\textrm{or $X_k$'s beam does not cover node $X_j$}
\end{array} \right.
\label{equ:interference}
\end{equation}
Fig. \ref{fig:interference-dir} shows that a transmission from node $X_k$ will not cause interference to $X_i$'s transmission since the antenna beam of $X_k$ does not cover receiver $X_j$.

We then derive the upper bounds on the per-node throughput capacity under Connectivity Condition, which is presented in Proposition \ref{prop:regime-i-DA}.

\begin{proposition}
\label{prop:regime-i-DA}
When Connectivity requirement dominates, the per-node throughput capacity contributed by ad hoc communications is $\lambda_a=O\big(\frac{4\pi^2}{\phi^2} \cdot \frac{nW_A}{H^3 \log^2 n}\big)$.
\end{proposition}

\textbf{Proof.} 
We first derive the probability that a node uses the ad hoc mode to transmit - $P(AH)=\pi H^2 r^2(n)$, 
which is the same as that of an \emph{MC-IS} network. 

Since each source generates $\lambda_a$ bits per second and there are totally $n$ sources, the total number of bits per second served by the whole network is required to be at least $n\cdot P(AH)\cdot \overline{h} \cdot \lambda_a$, where $\overline{h}$ is bounded by $\Theta(H)$, i.e., Lemma \ref{lemma:H} also holds for the case of Connectivity Condition of an \emph{MC-IS-DA} network. We next prove that $n\cdot P(AH)\cdot \overline{h} \cdot \lambda_a$ is bounded by $\frac{4\pi^2}{\phi^2} \cdot \frac{k_1}{\Delta^2 (r(n))^2} W_A$.

The maximum number of simultaneous transmissions on a particular channel denoted by $N_{\max}$ is affected by the number of interfering nodes in its neighborhood, which is determined by the size of the interference region. When we use directional antennas at both transmitter and receiver ends, the condition interference zone is $\frac{\theta^2}{(2\phi)^2}$ portion of that one when omni-directional antennas are used at both ends according to our extended interference model in Eq. (\ref{equ:interference}). Thus, $N_{\max}$ is upper bounded by $\frac{4\pi^2}{\phi^2} \cdot \frac{k_1}{\Delta^2 (r(n))^2}$, where $k_1>0$ is a constant, independent of $n$. Note that each transmission over the $\varpi$ channel is of $W_A/C_A$ bits/sec. Adding all the transmissions taking place at the same time over all the $C_A$ channels, we have the total number of transmissions in the whole network is no more than $\frac{k_1}{\Delta^2 (r(n))^2}\sum_{\varpi=1}^{C_A} \frac{W_A}{C_A} = \frac{k_1}{\Delta^2 (r(n))^2} W_A \cdot \frac{4\pi^2}{\phi^2} \textrm{bits/sec}$. Therefore, we have $n\cdot P(AH)\cdot \overline{h} \cdot \lambda_a \leq \frac{k_1}{\Delta^2 (r(n))^2} W_A \cdot \frac{4\pi^2}{\phi^2}$.

Combining the above results with Lemma \ref{lemma:H} yields $\lambda_a \leq \frac{k_1}{\Delta^2 r^2(n)} \cdot \frac{W_A}{n \pi H^3 r^2(n) } \leq \frac{k_2 W_A}{n H^3 r^2(n)} \cdot \frac{4\pi^2}{\phi^2}$, where $k_2$ is a constant. 

Besides, to guarantee that the network is connected with high probability (\textit{w.h.p.}), we require $r(n)>\sqrt{\frac{\log n}{\pi n}}$ \cite{Gupta:Kumar}. Thus, we have $\lambda_a \leq \frac{k_3 n W_A}{H^3 \log^2{n}}\cdot \frac{4\pi^2}{\phi^2}$, where $k_3$ is a constant.
\done

We then derive the upper bounds on the per-node throughput capacity under Interference Condition.

\begin{proposition}
\label{prop:regime-ii-DA}
When Interference requirement dominates, the per-node throughput capacity contributed by ad hoc communications is $\lambda_a=O\bigg(\frac{2\pi}{\phi} \cdot \frac{nW_A}{C_A^{\frac{1}{2}} H^3 \log^{\frac{3}{2}} n}\bigg)$.
\end{proposition}
\textbf{Proof.} 
The basic proving technique is the same as the proof of Proposition \ref{prop:regime-ii} as presented in Appendix A. To avoid repetitions, we only list the key steps in the derivations.

The main difference lies in the interference region of an \emph{MC-IS-DA} network, which is $\frac{\phi^2}{4\pi^2}$ of that one of an \emph{MC-IS} network \cite{Su:2003}. Therefore, we have 
\begin{displaymath}
\sum_{b=1}^{n \lambda_a P(AH)} \sum_{h=1}^{h(b)} \frac{\pi \Delta^2}{4} (r(b,h))^2 \cdot \frac{\phi^2}{4\pi^2} \leq W_A.
\end{displaymath}

This in-equation can be rewritten as
\begin{equation}
\label{eqn:th1-da}
\sum_{b=1}^{n \lambda_a P(AH)} \sum_{h=1}^{h(b)} \frac{1}{T_h} (r(b,h))^2 \leq \frac{4 W_A}{\pi \Delta^2 T_h} \cdot \frac{\phi^2}{4\pi^2}.
\end{equation}

Since the left hand side of this in-equation is convex, we have
\begin{equation}
\label{eqn:th2-da}
\Bigg(\sum_{b=1}^{n \lambda_a P(AH)} \sum_{h=1}^{h(b)} \frac{1}{T_h} r(b,h)\Bigg)^2 \leq \sum_{b=1}^{n \lambda_a P(AH)} \sum_{h=1}^{h(b)} \frac{1}{T_h} (r(b,h))^2.
\end{equation}

Joining (\ref{eqn:th1-da})(\ref{eqn:th2-da}), we have
\begin{displaymath}
\sum_{b=1}^{n \lambda_a P(AH)} \sum_{h=1}^{h(b)} r(b,h) \leq \sqrt{\frac{4 W_A T_h}{\pi \Delta^2 } \cdot \frac{\phi^2}{4\pi^2}}.
\end{displaymath}

From (\ref{eqn:th0}) in Appendix A, we have
\begin{displaymath}
\sum_{b=1}^{n \lambda_a P(AH)} \sum_{h=1}^{h(b)} r(b,h) \leq \frac{2\pi W_A}{\phi}\sqrt{\frac{2 n }{\pi \Delta^2 C_A}}.
\end{displaymath}

Besides, since $\lambda_a \cdot n \overline{l} \cdot P(AH) \leq \sum_{b=1}^{n \lambda_a P(AH)} \sum_{h=1}^{h(b)} r(b,h)$, we have
\begin{displaymath}
\lambda_a \leq \frac{\frac{2\pi W_A}{\phi}\sqrt{\frac{2 n}{\pi \Delta^2 C_A}}}{n \overline{l} \cdot P(AH)}
=\frac{\frac{2\pi W_A}{\phi}\sqrt{\frac{2 n}{\pi \Delta^2 C_A}}}{n \overline{h} r(n) \pi H^2 (r(n))^2} 
\leq \frac{\frac{2\pi W_A}{\phi}\sqrt{\frac{2}{\pi \Delta^2 n C_A}}}{\pi H^3 (r(n))^3}.
\end{displaymath}

Since $r(n) > \sqrt{\frac{\log n}{\pi n}}$, we finally have
$\lambda_a \leq \frac{k_4 n W_A}{C_A^{\frac{1}{2}} H^3 \log^{\frac{3}{2}}n} \cdot \frac{2\pi}{\phi}$.

Compared with the result of an \emph{MC-IS} network presented in Proposition \ref{prop:regime-ii}, an \emph{MC-IS-DA} network network has a capacity gain of $\frac{2\pi}{\phi}$. 
\done

Since the upper bounds on the throughput capacity under the destination-bottleneck condition and interface-bottleneck condition of an \emph{MC-IS-DA} are the same as those of an \emph{MC-IS} network, we ignore the proof here.
 
We next derive the lower bounds on the network capacity by constructing a network with the corresponding routing protocol and scheduling protocol with the satisfaction with each requirement. 

\emph{2. Lower bounds on Capacity of an MC-IS-DA Network}

Similar to the proof of lower bounds on the capacity of an \emph{MC-IS} network, we also have the following three steps: (1) cell construction, (2) designing routing scheme, (3) designing TDMA scheme to schedule the transmissions. Without repetitions, we only highlight the key steps in our proof.

(1) \emph{Cell Construction.} We divide the plane into $1/a(n)$ equal-sized cells, each of which is a square with area of $a(n)$. The cell size of $a(n)$ must be carefully chosen to fulfill the three requirements, i.e., the connectivity requirement, the interference requirement and the destination-bottleneck requirement. Differently,  we set $a(n)=\min\bigg\{\max\bigg\{\frac{100 \log n}{n}, \frac{ \log^{\frac{3}{2}}n}{C_A^{\frac{1}{2}} n}\cdot \frac{\phi}{2\pi}\bigg\}, \frac{\log^{\frac{3}{2}} n \cdot \log(H^2 \log n)}{n^{\frac{3}{2}} \cdot \log \log(H^2 \log n)}\bigg\}$ in our \emph{MC-IS-DA} network. Similarly, the maximum number of nodes in a cell can be upper bounded by Lemma \ref{lemma:no_nodes} in Section \ref{sec:lower-adhoc}.

The number of interfering cells around a cell in an \emph{MC-IS-DA} network is also bounded by a constant, which nonetheless is different from that of an an \emph{MC-IS} network. In particular, we have the following result.
\begin{lemma}
\label{lemma:interfering_cells-DA}
The number of cells that interfere with any given cell is bounded by a constant $k_9$ (where $k_9=81(2+\Delta)^2 \frac{\phi^2}{4\pi^2}$), which is independent of $a(n)$ and $n$. 
\end{lemma}
\textbf{Proof.} 
Suppose that there is a cell $D$ that can transmit with its 8 neighboring cells. The transmission range of each node in cell $D$, $r(n)$, is defined as the distance between the transmitter and the receiver. Since each cell has the size $a(n)$, $r(n)$ is no more than $3\sqrt{a(n)}$ (if including the cell itself, there are 9 cells). 

From the interference model, the transmission is successful only when the interfering nodes are $(1+\Delta)r(n)$ away from the receiver or the interfering nodes will not cause interference at the receiver (the beams of the interfering nodes do not cover the receiver). Let us consider that a transmitter $X_i$ within cell $B$ is transmitting a data packet to a receiver $X_j$ within cell $A$. Since the transmission range between $X_i$ and $X_j$ is $r(n)$, the distance between two transmitter $X_k$ and $X_i$ must be less than $(2+\Delta)r(n)$, if $X_k$ causes the interference with $X_j$. Thus, an interfering area is loosely bounded within a square with an edge length of $3(2+\Delta)r(n)$. 

Meanwhile, to ensure a successful transmission, the beams of the two nodes are pointed at each other. Thus, only the nodes within the receiving beam of $X_j$ can interfere with the reception at $X_j$. Besides, only when a transmitter adjusts its beam to the receiver, it can interfere with the receiver. So, the interfering probability is $(\frac{\phi}{2\pi})^2$. Combining the two observations, there are at most $k_9=\frac{(3(2+\Delta)r(n))^2}{a(n)}\cdot (\frac{\phi}{2\pi})^2=81(2+\Delta)^2 \frac{\phi^2}{4\pi^2}$ interfering cells. Hence, the number of interfering cells is bounded by $81(2+\Delta)^2 \frac{\phi^2}{4\pi^2}$, which is a constant $k_9$ independent of $a(n)$ and $n$.
\done

(2) \emph{Routing Scheme.} The routing scheme is the same as that of an \emph{MC-IS} network. So, we ignore the detailed proof here.

(3) \emph{Scheduling Transmissions.}
We next design a scheduling scheme to transmit the traffic flows assigned in a \textit{routing scheme}. We propose a TDMA scheme to schedule transmissions. In this scheme, one second is divided into a number of \textit{edge-color} slots and at most one transmission/reception is scheduled at every node during each edge-color slot. Each edge-color slot can be further split into smaller \textit{mini-slots}. The only difference lies in the mini-slot. In particular, every cell has at most a constant number of interfering cells with a factor $(\frac{\phi}{2\pi})^2$, and each cell has $\Theta(n a(n))$ nodes. Thus, each node has at most $O((\frac{\phi}{2\pi})^2 n a(n))$ edges in the interference graph. It is shown that a graph of degree at most $k$ can be vertex-colored with at most $k+1$ colors \cite{Douglas:2001}. Hence, the interference graph can be vertex-colored with at most $O((\frac{\phi}{2\pi})^2 n a(n))$ colors. Then, we use $k_{10} (\frac{\phi}{2\pi})^2 n a(n)$ to denote the number of vertex-colors (where $k_{10}$ is a constant).

We need to schedule the interfering nodes either on different channels, or at different mini-slots on the same channel since two nodes assigned the same vertex-color do not interfere with each other, while two nodes stained with different colors may interfere with each other. We divide each edge-color slot into $\left\lceil \frac{k_{10} n a(n)}{C_A} \cdot \frac{\phi^2}{4\pi^2} \right\rceil$ mini-slots on every channel, and assign the mini-slots on each channel from 1 to $\left\lceil \frac{k_{10} n a(n)}{C_A} \cdot \frac{\phi^2}{4\pi^2}\right\rceil$. A node assigned with a color $s$, $1\leq s \leq  k_{10} n a(n) \cdot \frac{\phi^2}{4\pi^2}$, is allowed to transmit in mini-slot $\left\lceil \frac{s}{C_A} \right\rceil$ on channel $(s \textrm{ mod } C_A) +1$.

We next prove the constructive lower bounds of the capacity. 
\begin{proposition}
\label{prop:lower-adhoc-da}
The achievable per-node throughput $\lambda_a$ contributed by ad hoc communications is as follows.
\begin{enumerate}
\item[1)] When Connectivity requirement dominates, $\lambda_a$ is $\Omega\Big(\frac{nW_A}{H^3 \log^2 n } \cdot \frac{4\pi^2}{\phi^2} \Big)$ bits/sec;
\item[2)] When Interference requirement dominates, $\lambda_a$ is $\Omega\Big(\frac{nW_A}{H^3 C_A^{\frac{1}{2}}\log^{\frac{3}{2}}n} \cdot \frac{2\pi}{\phi}\Big)$ bits/sec;
\item[3)] When Destination-bottleneck requirement dominates, $\lambda_a$ is $\Omega\bigg(\frac{n^{\frac{3}{2}} \log \log (H^2 \log n) W_A}{C_A H^3 \log^{\frac{3}{2}}n \cdot \log(H^2 \log n)}\bigg)$ bits/sec;
\item[4)] When Interface-bottleneck requirement dominates, $\lambda_a$ is $\Omega\big(\frac{W_A}{C_A}\big)$.
\end{enumerate}
\end{proposition}
\textbf{Proof.}
Since each edge-color slot with a length of $\Omega\big(\frac{1}{H^3 a(n)} \cdot \frac{\phi^2}{4\pi^2}\big)$ seconds is divided into $\left\lceil \frac{k_{10} n a(n)}{C_A}\right\rceil$ mini-slots over every channel, each mini-slot has a length of $\Omega\Big(\big(\frac{1}{H^3 a(n)}\big)/\left\lceil \frac{k_{10} n a(n)}{C_A}\cdot \frac{\phi^2}{4\pi^2}\right\rceil\Big)$ seconds. Besides, each channel can transmit at the rate of $\frac{W_A}{C_A}$ bits/sec, in each mini-slot, $\lambda_a=\Omega\Big(\frac{W_A}{C_A H^3 a(n) \cdot \left\lceil \frac{k_{10} n a(n)}{C_A}\cdot \frac{\phi^2}{4\pi^2}\right\rceil}\Big)$ bits can be transported. Since $\left\lceil \frac{k_{10} n a(n)}{C_A}\cdot \frac{\phi^2}{4\pi^2}\right\rceil \leq \frac{k_{10} n a(n)}{C_A} \cdot \frac{\phi^2}{4\pi^2} +1$, we have $\lambda_a=\Omega\Big(\frac{W_A}{k_{10} H^3 a^2(n) n \cdot \frac{\phi^2}{4\pi^2} + H^3 a(n) C_A }\Big)$ bits/sec. Thus, $\lambda_a=\Omega\Big(MIN_O\Big(\frac{W_A}{H^3 a^2(n) n\cdot \frac{\phi^2}{4\pi^2}},\frac{W_A}{H^3 a(n) C_A}\Big)\Big)$ bits/sec. Since $a(n)$ is set to $\min\Big\{\max\Big\{\frac{100 \log n}{n}, \frac{ \log^{\frac{3}{2}}n}{C_A^{\frac{1}{2}} n} \cdot \frac{\phi}{2\pi}\Big\}, \frac{\log^{\frac{3}{2}} n \cdot \log(H^2 \log n)}{n^{\frac{3}{2}} \cdot \log \log(H^2 \log n)}\Big\}$. Substituting the three values to $\lambda_a$, we have the results 1), 2) and 3). Besides, each interface can transmit or receive at the rate of $\frac{W_A}{C_A}$ bits/sec. Thus, $\lambda_a=\Omega\big(\frac{W_A}{C_A}\big)$, which is the result 4).

Following the similar proving steps in Section \ref{sec:lower-adhoc} and Section \ref{sec:infra}, we can finally obtain Corollary \ref{cor:throughput}.
\done
\bibliography{IEEEabrv,mcisref}

\end{document}